\documentclass[10pt, conference, letterpaper]{IEEEtran}
\usepackage{amsmath}
\usepackage{amssymb}
\usepackage{amsthm}
\usepackage{tikz}
\usepackage{pgfplots}
\usepackage{multirow}
\usepackage{graphicx}
\usepackage{bbm}
\usepackage{breqn}
\usepackage{lipsum}
\usepackage[ruled,linesnumbered,noend,noline]{algorithm2e}
\usepackage{mathtools}
\usepackage{hyperref}
\newcommand\numberthis{\addtocounter{equation}{1}\tag{\theequation}}
\makeatletter
\def\BState{\State\hskip-\ALG@thistlm}
\makeatother

\ifCLASSINFOpdf
\else
\fi
\tikzset{
	state/.style={
		rectangle,
		rounded corners,
		draw=black, very thick,
		minimum height=2em,
		inner sep=2pt,
		text centered,
	},
}
\begin{document}
\bibliographystyle{plain}

\title{\emph{InferBeam}: A Fast Beam Alignment Protocol for Millimeter-wave Networking}
\author{Sai Qian Zhang, H.T. Kung and Youngjune Gwon\\
	\IEEEauthorblockA{John A. Paulson School of Engineering and Applied Sciences\\
		Harvard University}}
\maketitle
\begin{abstract}

We introduce fast millimeter-wave base station (BS) and its antenna sector selection for user equipment based on its location. Using a conditional random field inference model with specially designed parameters, which are robust to change of environment, InferBeam allows the use of measurement samples on best beam selection at a small number of locations to infer the rest dynamically. Compared to beam-sweeping based approaches in the literature, InferBeam can drastically reduce the setup cost for beam alignment for a new environment, and also the latency in acquiring a new beam under intermittent blockage. We have evaluated InferBeam using a discrete event simulation. Our results indicate that the system can make best beam selection for $98\%$ of locations in test environments comprising small-sized apartment or office spaces, while sampling fewer than $1\%$ of locations. InferBeam is a complete protocol for best beam inference that can be integrated into millimeter-wave standards for accelerating the much-needed fast and economic beam alignment capability.
\end{abstract}

\section{Introduction}
The forthcoming 5th generation (5G) mobile networks are expected to embrace the underutilized millimeter-wave (mm-wave) frequencies at 28\,GHz, 38\,GHz, or higher to deliver high data rates up to 10\,Gbps over wireless. According to a white paper by Cisco \cite{ciscowhitepaper}, the global wireless data traffic has increased 18-folds over the past five years. At current rate, wireless data growth will lead to exhaustion of spectrum resources which are mainly situated in the lower frequencies (e.g. below 6GHz). Millimeter-wave frequencies provide an opportunity to mitigate global spectrum shortage---in particular, the 60-GHz mm-wave band with up to 9-GHz unlicensed bandwidth has been identified as one of the key enablers for the high-speed data transmission over 5G.

In fact, the utility of the mm-wave bands is roughly equivalent to 200 times more bandwidth than the allocation of today's Wi-Fi and cellular networks. Multi-Gbps data rates will enable a new class of applications such as Wireless data center, Tetherless VR/AR, Information showers, and Wireless chip interconnection \cite{mmwaveusecase}. These are likely to make mm-waves a central proponent of the 5G cellular standard. In addition, mm-waves have also been pursued in other wireless communication standards such as the IEEE 802.11ad for Wi-Fi and IEEE 802.15.3c for wireless personal area network (PAN) \cite{802.11ad, 802.15.3c}.    

However, the mm-waves is known to suffer greater attenuation. A simple calculation by using the Friis's formula indicates that mm-wave at 60 GHz decays 21.6 dBm more than 5 GHz and 28 dBm more than 2.4-GHz. To overcome this high-attenuation problem, the IEEE 802.11ad \cite{802.11ad}, for example, defines a communication scheme for highly directional antennas. The scheme takes advantages of antenna gain through beamforming to compensate greater attenuation of the mm-waves. The IEEE 802.11ad discretizes the search space for beamforming by dividing the antenna azimuth into \emph{virtual sectors}. These sectors can be implemented with either phase array antenna or multiple directional antenna elements. With a high directionality of mm-wave, the communication between the transmitter and the receiver can achieve the highest data rate only when their beams are aligned, causing a highly directional signal focus. The current scheme for the IEEE 802.11ad requires that the transmitting antenna scan the whole space exhaustively to find the best beam steering direction that will generate the largest receiver SNR through the ACK frame from the receiver. This process, however, may take up to several seconds, which can seriously degrade the system performance. For wide adoption, the latency problem in beam alignment at
mm-wave frequencies must be addressed.

Moreover, the antenna selection problem still exists in the 60Hz frequency band. To provide individual devices with exceedingly high bandwidth, the next-generation wireless networks will not rely only on intermittently placed cell towers, but also on locally deployed, overlapping, dense “small cells”, each of which contains its own BS. Devices connecting to the network will have to select which antenna will provide optimal performance. Consider several BSs in the 3D space, each BS broadcasts the beacons through its every sector. The user equipment (UE) will discover the BSs by receiving the beacons from each BS, then select the BS to associate with based on the SNR. However, the BS discovery latency will be as high as 12.9s \cite{linkstateprofile} due to the high directionality of mm-wave. 

In this paper, we propose InferBeam for inferring millimeter-wave BS and aligning beams based on conditional random field (CRF) \cite{patternml}, a class of statistical modeling method for structured inference in machine learning. CRF is applied popularly to computer vision tasks such as image segmentation and object recognition. Our key idea is to train CRFs to segment a wireless environment modeled as a 3D grid such that via CRF inference every discrete point within the environment is associated with its optimal antenna and sector. CRF is used to reflect spatial correlation on antenna and sector selection of neighboring points in the 3D space, allowing the model to incorporate domain-specific knowledge (e.g. location of obstacles) into the segmentation and improving the overall performance of the technique.

InferBeam is differentiated from a typical mm-wave beamforming method such as the IEEE 802.11ad that relies on exhaustive search of optimal beamforming direction. Such method will incur a high overhead problematic in the 5G networking setting wherein mm-wave BSs are required to accommodate many mobile users. We have designed the InferBeam protocol and implemented the CRF inference modules for evaluation. Our initial results are promising that InferBeam can make best beam selections for $98\%$ of locations in the test environment which comprise small-sized apartment or office spaces, while sampling fewer than $1\%$ of locations.

The rest of this paper is organized as follows. In the next section, we will present the background and state the motivation. Section III describes the learning and training for CRF. The InferBeam system and protocol design will be presented in Section IV. Our empirical evaluation follows in Section V, the paper concludes with Section VI. 
\section{Background and Motivation}
\begin{figure}
	\centerline{\includegraphics[width=8.2cm,height=3.8cm]{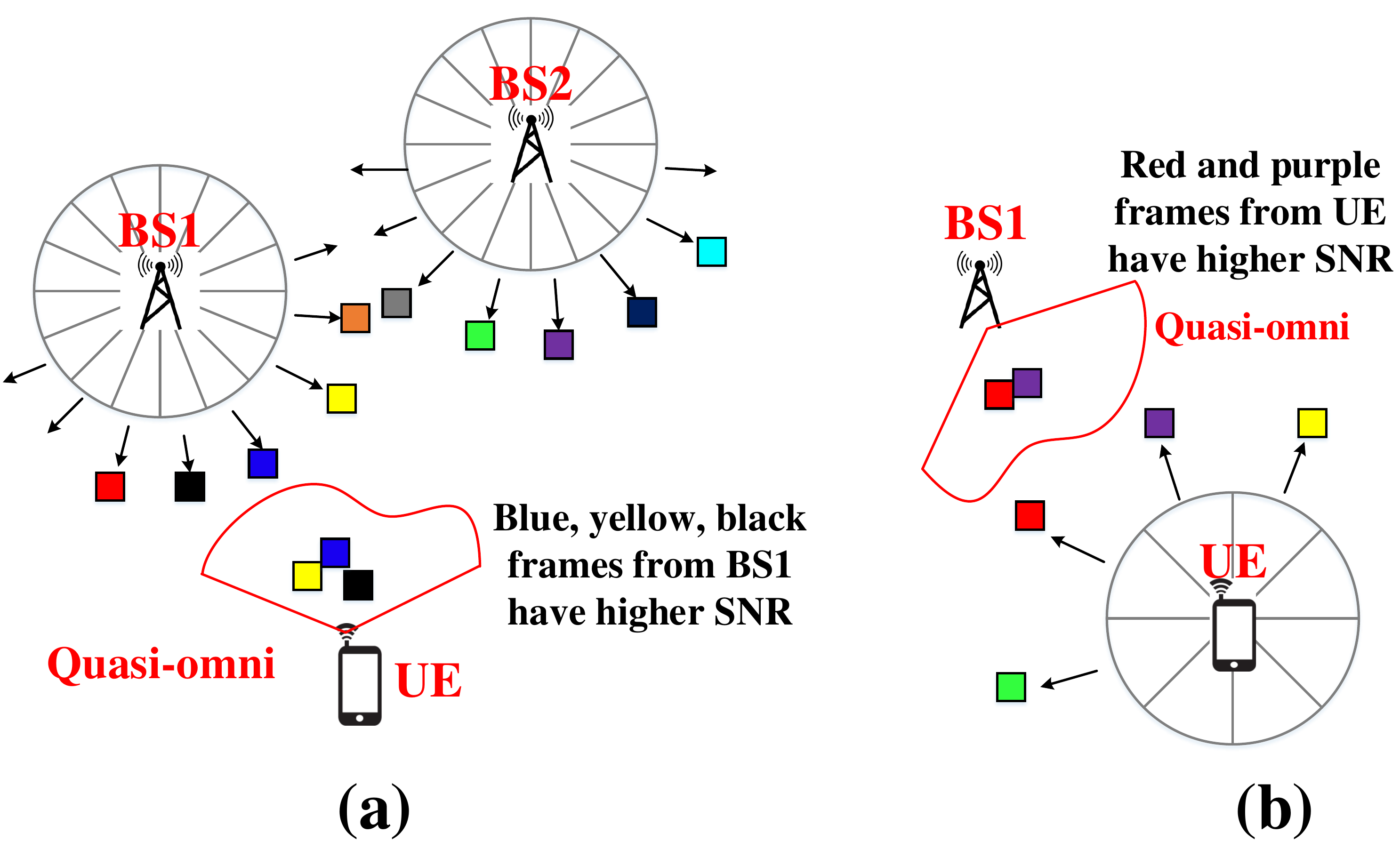}}
	\caption[U-example]{Sector Level Sweep phase}
	\label{fig:sls}
\end{figure}
To motivate our work, we provide background on existing mm-wave beamforming techniques. We also review the basics of conditional random field and discuss related work.
\subsection{IEEE 802.11ad Beamforming for mm-waves}
Beamforming is a technique used in achieving directional transmission or reception of wireless signals. Typically, the directivity is created through the adjustment of multiple antenna elements. Beamforming procedure defined by IEEE 802.11ad can take place in two phases. First, coarse-grain antenna sector selection needs to be performed by sector-level sweep (SLS). Fig. \ref{fig:sls}(a) and \ref{fig:sls}(b) illustrates SLS. Each BS transmits probe frames at different sectors with unique identifiers. UE receives these frames from both BSs with a quasi-omni-directional antenna pattern, selecting the BS and a its coarse-grained BS transmit sector which has higher SNR (Fig. \ref{fig:sls}(a)), and sending the best transmit sector ID to that BS. The same procedure is inverted to find a list of transmit sectors of UE (Fig. \ref{fig:sls}(b)). This procedure generates high overhead since each probe frame is transmitted from each sector at the lowest PHY rate (MCS 0), which may take up to several seconds. 

Given an antenna sector, beam refinement phase (BRP) follows. During BRP, the coarse-grained transmit sectors founded in SLS phase is refined. The optimum transmit and receive sectors are determined to optimize the data rate between BS and UE. In fact, BRP is faster than SLS. The frame exchanges in BRP can rely on the directional link built during the SLS phase. 

\subsection{Conditional random field (CRF)}
Conditional random field (CRF) is a probabilistic learning framework for labeling and segmenting structured data such as grid spaces, trees, and lattices. In machine learning, a label (or ground-truth mark) serves as a guide to train a learning unit. CRF models a conditional probability distribution over a sequence of labels given a sequence of observations, rather than inferring the joint distribution, which is a much harder learning task. Assume a graphical model $G(V,E)$, where each node $v$ in the grid is associated with a random variable $x_{v}\in X$ which represents the label assigned to this node, and $X$ is the set of all possible labels. Furthermore, each node and edge are associated with a \emph{node potential} and \emph{edge potential}. In reality, adjacent nodes tend to have the same label selection. We capture this tendency using edge potential functions. In addition to this, we have certain “measurements” for the nodes which also influence our belief about their true values. These measurements are represented by node potentials. By convention, we use $\phi_{v}(x_{v})$ and $\psi_{vv'}(x_{v},x_{v'}),$ to denote the node potential for $v\in V$ and edge potential for $(v,v')\in E$ respectively. Let $x_{V}=\{x_{v\in V}\}$ to represent the label selections for all the nodes in the grid. 
The joint probability $p(x_{V})$ is defined as the product of all the node potentials and edge potentials:
\begin{equation}
P(x_V)=\frac{1}{Z}\prod_{v\in V} \phi_{v}(x_{v})\prod_{(v,v')\in E} \psi_{vv'}(x_{v},x_{v'})
\end{equation} 
where $Z$ is the normalization constant. 
\subsection{Motivation}
Recognizing the drawbacks on the BS selection and beamforming procedures eminent in the mm-wave standards (e.g., 802.11ad), we are motivated to develop a fast beamforming approach for mm-waves. In our illustration, we assume that the optimal antenna and sector selections are known for the two samples, UE1 and UE2. We further assume that UE3 is located between UE1 and UE2, the distance between UE1 and UE3 is 0.5m, while the distance between UE2 and UE3 is 0.7m. To infer the antenna and sector selection for UE3, we may take the information from both samples into consideration. Moreover, we need to find the spatial correlation on BS and sector selection between the sample points and the point we want to infer. In this paper, we use CRF to construct an undirected graphical model that captures a wireless environment.
This motivates the InferBeam, which uses CRF to infer optimal BS selection and antenna sector decision, using only small number of samples from the environment's 3D grid. With offline bootstrapping, our proposed InferBeam can provide a low-latency decision for UE's optimal beamforming configuration. InferBeam has the following advantages: 1. InferBeam enables quick beamforming by skipping the SLS phase. 2. InferBeam enables fast beam adjustment as environment changing (i.e. blockage).
\subsection{Related work}
The existing literatures have studied a number of unique properties of the mm-wave, including the increased signal attenuation of propagation, high sensitivity to the obstacle blockage and transmitting/receiving device motion \cite{mmmeasure1} \cite{mmmeasure} \cite{linkstateprofile}. Furthermore, The authors of \cite{linkstateprofile} show that unlike the legacy Wi-Fi, even if the transmitter is sending the frames with omni-directional pattern, the discovery latency for the frames to reach receiver is heavily depended on the location and mobility of the receiver. The receiver's angular receiver signal strength is sparsely distributed, implying that the receiver can only receive the frames from very limited number of angles. Several papers \cite{mmmeasure1}\cite{mmmeasure4}\cite{mmmeasure5} have provided some precise measurements concerning mm-wave propagation in indoor/outdoor environments. More specifically, the authors of \cite{mmmeasure1} have presented a directional and omni-directional path loss model for both LOS and NLOS environments.

Several beamforming techniques have been proposed to reduce the beamforming latency. In \cite{blindsteering}, the authors present a system called "Blind Beam Steering" which utilizes the reading from the legacy Wi-Fi to infer the beamforming direction of mm-wave. A set of codebook based techniques for beamforming are proposed in \cite{codebook1}\cite{codebook2}, where the basic idea is to assign each beam angle a unique code so that multiple beam angles can be tested simultaneously, thereby accelerating the training process. In AgileLink \cite{agilelink}, fast beamforming is achieved through hashing the beam directions with a few selected hashing functions, and then identifying the correct alignment by analyzing the change on energy for different hashing functions. In \cite{wififingerprint}, correlation is calculated offline between the fingerprint of UE Wi-Fi signal and mm-wave best beam ID, and then given a new UE with its Wi-Fi fingerprint, the optimal beam direction can be found. The authors of \cite{beamspy} design a system called "Beamspy" which can predict the quality of all the mm-wave beams under the blockage so that the beam can be re-aligned rapidly. Departing from these prior approaches, the proposed InferBeam protocol in this paper focuses on minimizing the required samples in inferring best beams.
\section{CRF Training and Inference}
In this section, we will describe our CRF inference for best beam inference, and how the CRF training is performed.
\subsection{CRF Inference}
First we give a mathematical analysis on CRF inference. More specifically, we are interested in solving such a problem: given a graphical model $G=(V,E)$ which is represented by a $a\times b\times c$ 3D grid, its model parameter $\theta$, and ground truth labels $x^{*}_{S}$ of some nodes $S\subseteq V$, estimate the marginal distributions for the rest of nodes in the grid.
Before giving the expressions on node potential and edge potential, we give a definition for \emph{physical-hops (p-hops)}. Given a node $v_{0}$ in the 3D grid, rank the rest of the nodes in the grid based on its line-of-sight distance to $v_{0}$. The rank of a node $v$ is defined as the distance in \emph{physical-hops} between $v$ and $v_{0}$. An example is shown in Fig. \ref{fig:2dgrid}, for the red node in the center, the blue, yellow, purple, green, black nodes are 1,2,3,4,5 p-hops away from it respectively. 

Next, we give the expressions for node potential and edge potential. Denote $X$ the set of all possible labels, and $x_{v} \in X$ the label of node $v$. The node potential and edge potential are defined as follows:
\begin{equation}
\phi_{v}(x_{v})=exp(\sum_{k=0}^{K} w_{k} \sum_{s\in S,d(s,v)=k} \mathbbm{1}_{x^{*}_{s}=x_{v}})
\end{equation} 
\begin{equation}
\psi_{vv'}(x_{v},x_{v'})=exp(m_{vv'}\mathbbm{1}_{x_{v}\neq x_{v'}})
\end{equation} 
$w_{k}$ is a measure of spatial correlation on the label selection between a node and the nodes $k$ p-hops away from it, and $K$ is the range of distances in p-hops.  When $k=0$, $w_k$ will be infinity ideally. $S$ is set of sample nodes taken at some fixed locations in the grid and $x^{*}_{s}$ is the ground-truth label of $s$. $d(s,v)$ is the distance in p-hops between $s$ and $v$. $m_{vv'}$ is a parameter which measures the degree of penalty for neighboring nodes that are assigned different labels. By the definition of edge potential, we have $\psi_{vv'}(x_{v},x_{v'}) = exp(m_{vv'})$ if $x_{v}\neq x_{v'}$ and $\psi_{vv'}(x_{v},x_{v'}) = 1$ otherwise. By making $m_{vv'}$ small, we require the high consistency on label selections between the neighboring nodes. The node potential $\phi(x_{v})$ considers the sample nodes which are within $K$ p-hops from $v$, calculate the weighted sum of number of samples which have the same label as $v$, and then take the exponential of it to make the node potential positive. Substituting $(2)$ and $(3)$ into $(1)$, we have $p(x_{V})$, the joint probability of label selections:
\begin{multline}
P(x_{V})=\\
\frac{1}{Z}\prod_{v\in V} exp(\smashoperator{\sum_{k=0}^{K}} w_{k} \smashoperator{\sum_{\substack{s\in S\\ d(s,v)=k}}} \mathbbm{1}_{x^{*}_{s}=x_{v}}) \smashoperator{\prod_{(v,v')\in E}}exp(m_{vv'}\mathbbm{1}_{x_{v}\neq x_{v'}})
\end{multline} 
and the marginal distribution $P(x_{v})$ can be found by summing $P(x_{V})$ over the other variables $x_{v\backslash V}$:
\begin{equation}
P(x_{v}) = \sum_{v'\in {V\backslash \{v\}}}\sum_{x_{v'}} P(x_{V})
\end{equation}

\subsection{CRF Learning}
Next, we discuss the training procedure for the CRF. The objective is to find the parameters $\theta=\{w_{k},m_{vv'}\}$, $ 1\leq k\leq K$ and $(v,v')\in E$, given the training set $D=\{D_{1},D_{2},...,D_{R}\}$. $D_{r} = \{x^{r}_{v\in V}\}$ is a $a\times b\times c$ matrix which contains the optimal labels for the 3D grid.
We are not interested in $w_{0}$ since $w_0$ equals infinity theoretically. By Baysian's rule, we have: 
\begin{align}
&\prod_{r=1}^{R} P(D_{r}|\theta) P(\theta) = P(D|\theta)P(\theta)\propto P(\theta|D)
\end{align}
where $P(\theta)$ is the prior distribution of $\theta$. The prior distribution represents our initial belief in the values of $\theta$.
Assume the prior distributions of $w_k$ and $m_{vv'}$ follows Gaussian, a very popular choice for the informative prior distribution, with mean $\mu_{w_{k}}$, $\mu_{m}$ and standard deviation $\sigma_{w_{k}}$, $\sigma_{m}$:
\begin{equation}
P(\theta) = \prod_{k=1}^{K} \mathcal{N}(w_{k};\mu_{w_{k}},\sigma_{w_{k}})\prod_{(v,v')\in E}\mathcal{N}(m_{vv'};\mu_{m},\sigma_{m})\numberthis 
\end{equation}
where $\mathcal{N}(x;\mu,\sigma)$ represents the Gaussian pdf with variable $x$, mean $\mu$ and standard deviation $\sigma$. Applying $(4)$ and $(10)$ to $(9)$, we have $\frac{1}{R}log P(\theta|D)=\frac{1}{R}\sum_{r=1}^{R}\log P(D_{r}|\theta) + \frac{1}{R}\log  P(\theta)+ C$. After substituting the expressions for $P(D|\theta)$ and $P(\theta)$, we have:
\begingroup\makeatletter\def\f@size{9}\check@mathfonts
\begin{align*}
&\frac{1}{R}log P(\theta|D) = -logZ(\theta) + \frac{1}{R}\sum_{r=1}^{R}\bigg(\sum_{v\in V}\sum_{k=0}^{K}w_{k}\smashoperator{\sum_{\{s|d(s,v)=k\}}} \mathbbm{1}_{x^{r}_{s}=x^{r}_{v}} + \\& \smashoperator{\sum_{(v,v')\in E}} m_{vv'} \mathbbm{1}_{x^{r}_{v}\neq x^{r}_{v'}} \bigg)+ \sum_{k=1}^{K} -\frac{(w_{k}-\mu_{w_{k}})^{2}}{2R\sigma_{w_{k}}^{2}} +  \smashoperator{\sum_{(v,v')\in E}} -\frac{(m_{vv'}-\mu_{m})^{2}}{2R\sigma_{m}^{2}} \\&+ C \numberthis 
\end{align*}
\endgroup
where $Z(\theta)$ is the normalization constant for $p(x_{V};\theta)$, $x^{r}_{v}$,$x^{r}_{s}$ is label of $v$ and $s$ in $D_{r}$ and $C$ is the terms which are not related to $\theta$. Given the expression for the posterior distribution, we want to find $w_{k}, m_{vv'}$ which maximize $(11)$. We have the following theorem for the gradient:
\newtheorem{mydef1}{Theorem}
\begin{mydef1}
	We have the following equations for the gradient: \\
	$\frac{\partial \frac{1}{R}log P(\theta|D)}{\partial w_{k}}= E_{D}(\sum_{v\in V} \sum_{\{s|d(s,v)=k\}} \mathbbm{1}_{x^{*}_{s}=x_{v}})-E_{P(x_{V};\theta)}(\sum_{v\in V}\sum_{\{s|d(s,v)=k\}} \mathbbm{1}_{x^{*}_{s}=x_{v}}) + \frac{\mu_{w_{k}}-{w}_{k}}{R\sigma_{w_{k}}^{2}}$ and  $\frac{\partial \frac{1}{R}log P(\theta|D)}{\partial m_{vv'}}= E_{D}(\mathbbm{1}_{x_{v}\neq x_{v'}}) - E_{P(x_{V};\theta)} (\mathbbm{1}_{x_{v}\neq x_{v'}})+\frac{\mu_{m}-m_{vv'}}{R\sigma_{m}^{2}}$, where $E_{D}(.)$ is the empirical expectation over the training data $D$ and $E_{P(x_{V};\theta)}(.)$ is the expectation over $P(x_{V};\theta)$. 
\end{mydef1}
\begin{proof}
	See appendix B.
\end{proof}
After getting the gradient of $\frac{1}{R}log P(\theta|D)$, we can use gradient ascent method to find the optimal $\theta$, as described in Algorithm 2. 
\begin{algorithm}
	\footnotesize
	\textbf{Input}: $\theta^{(0)}=\{w_{k}^{(0)},m_{vv'}^{(0)}\},\forall 1\leq k\leq K, (v,v')\in E$, step size $\eta$, stopping criterion $\Delta$, maximum number of iteration $I_{max}$, training data $D$ \\
	\textbf{Output}: $\theta^{*}=\{w_{k}^{*},m_{vv'}^{*}\},\forall 1\leq k\leq K, (v,v')\in E$ \\
	\For{i=1,...,$I_{max}$}{ 
		Compute $\frac{\partial \frac{1}{R}\log P(\theta|D)}{\partial w_{k}}$,$1\leq k\leq K$ and $\frac{\partial \frac{1}{R}\log P(\theta|D)}{\partial m_{vv'}}$, $(v,v')\in E$ with equations of gradients in Theorem 2.\\
		Compute $w_{k}^{(i)}=w_{k}^{(i-1)}+ \eta\frac{\partial \frac{1}{R}\log P(\theta|D)}{\partial w_{k}}$,$1\leq k\leq K$ and $m_{vv'}^{(i)}=m_{vv'}^{(i-1)}+ \eta\frac{\partial \frac{1}{R}\log P(\theta|D)}{\partial m_{vv'}}$, $(v,v')\in E$\\
		\If{$||w_{k}^{(i)}-w_{k}^{(i-1)}||\leq \Delta$ and $||m_{vv'}^{(i)}-m_{vv'}^{(i-1)}||\leq \Delta$ $\forall k,(v,v')$}{
			$m_{vv'}^{*} = m_{vv'}^{i}$ and $w_{k}^{*} = w_{k}^{i}$ $\forall$ $k$, $(v,v')$ \\
			\textbf{Break};
		}
	}
	\textbf{Return} $\theta^{*}$
	\caption{Gradient Ascent Algorithm}
\end{algorithm}
Furthermore, Theorem 3 states that $\theta^{*}$ returned by Algorithm 2 is the global optimum of $(11)$.  
\begin{mydef1}
	As $\Delta \rightarrow 0$, the $\theta^{*}$ returned by Algorithm 2 approaches the global maximum of $P(\theta|D)$.
\end{mydef1}
\begin{proof}
	See Appendix C.
\end{proof}

\section{AutoBeam System and Protocol Design}
In this section, we describe the InferBeam system and protocol in detail. The InferBeam system consists of two cascaded CRFs, CRF1 and CRF2. CRF1 is dedicated for BS selections while CRF2 will be used to select transmit sector.

\subsection{Inference}
A beam selection $v\in V$ is a tuple $(BS\_ID_{v}, Sec\_BS\_ID_{v}, Sec\_UE\_ID_{v})$, containing the BS ID, BS and UE transmit sector IDs. Each $v$ is associated with an identifier $Beam\_ID_{v}$. Hence, the beam tuple can describe a beamforming connection between BS and UE. We also assign a $Sec\_ID_{v}$ to the sub-tuple $(Sec\_BS\_ID_{v}, Sec\_UE\_ID_{v})$. Assigning $Sec\_BS\_ID$ and $Sec\_UE\_ID$ is based on the physical locations of the sectors. The sector with the largest sector ID and smallest sector ID are next to each other because of the circularly placement. We use notation $x^{*}_{S}$ for an optimal beam selection in a sample set $S$, containing the optimal beam selection for the given location $\{(BS\_ID^{*}_{s}, Sec\_BS\_ID^{*}_{s}, Sec\_UE\_ID^{*}_{s})\}$, $\forall s\in S$. 

The output of CRF inference includes $\{Q(x_{v})\}$, $\forall v \in V, x_{v}\in X$, where for every $v\in V$ in the grid, CRF returns a $|X|\times 1$ probability vector $\vec{Q}_{v}=[Q(x_{v}=x)]^\top, x\in X$. That is, for each point $v\in V$ in the grid, CRF returns a probability vector for each beam selection. We define \emph{beam selection map} $B_v$ for each $v$. Each entry of $B_{v}$ is a four-tuple which consists of $BS\_ID_{v}$, $Sec\_BS\_ID_{v}$, $Sec\_UE\_ID_{v}$, and its probability. Sorting $B_{v}$ in a descending order of the probability measure, we obtain $B^{sorted}_{v}$. An example of sorted $B_{v}$ is shown in Fig.~ \ref{fig:online_example}(a). We can also aggregate all the $B_{v}, v\in V$ to get $B = \{B_{v},v\in V\}$. Denote $X^{BS}$ and $X^{Sec}$ the set of BS IDs and sector tuple IDs, and assume all the BSs have same number of transmit sectors, and so do the UEs. The Backend Inference Algorithm (BIA) is presented in Algorithm 3.
\begin{algorithm}
	\footnotesize
	\textbf{Input}: Set of samples $S$, optimal beam selections $x_{S}^{*}$, CRF1, CRF2 \\
	\textbf{Output}: $B=\{B^{sorted}_{v}\}$,for $v\in V$ \\
	\For{$v\in V$}{
		Use $x^{BS*}_{S}$ and CRF1 as inputs of Algorithm 1 to calculate the probability vector $\vec{P}^{BS}_{v}=[Q(BS\_ID_{v}=x)]^\top$, $x\in X^{BS}$ 
	}
	\For{$x\in X^{BS}$}{
		\For{$v\in V$}{
			Denote $S_{x}$ the set of samples which has $BS\_ID^{*}=x$, use $x^{Sec*}_{S_{x}}$ and CRF2 as input of Algorithm 1 to calculate the probability vector $\vec{P}^{Sec}_{v,BS\_ID^{*}=x}=[Q(Sec\_ID_{v}=y|BS\_ID_{v}^{*}=x)]^\top, y\in X^{Sec}$
		}
	}
	\For{$v\in V$}{ 
		\For{each $x\in X^{BS}$}{
			\For{each $y\in X^{Sec}$}{
				Calculate $P(BS\_ID_{v}=x,Sec\_ID_{v}=y)=P(x,y) = P(y|x)P(x) = \vec{P}^{Sec}_{v,BS\_ID_{v}^{*}=x}(y)\vec{P}^{BS}_{v}(x)$, where $\vec{P}^{Sec}_{v,BS\_ID_{v}^{*}=x}(y)$ and $\vec{P}^{BS}_{v}(x)$ is the yth and xth element of $\vec{P}^{Sec}_{v,BS\_ID_{v}^{*}=x}$ and $\vec{P}^{BS}_{v}$\\
				Append $(x,y,P(x,y))$ to $B_{v}$ \\
			}		
		}	
		Sort $B_{v}$ in descending order based on the probability of each entry\\
	}
	
	\textbf{Return} $B^{sorted}=\{B^{sorted}_{v},v\in V\}$
	\caption{\textbf{Backend Inference Algorithm (BIA)}}
\end{algorithm}
BIA uses $x^{BS*}_{S}$ and CRF1 to infer the BS selection for the rest of the points in the grid (lines 3--4). For every $x$ in the $X^{BS}$, BIA picks the samples from $S$ whose $BS\_ID^{*}_{s}$ equals $x$, using their sector tuples and CRF2 to infer the sector tuples the rest of the points in the grid (lines 5--7). BIA then uses the probability vectors returned by CRF1 and CRF2 to calculate the joint probability of BS and sector tuple selection (line 11), generating the beam selection map for each $v$ (line 12). $B_{v}$ is then sorted in a descending order of the probability measure (line 13). Finally, BIA returns the sorted beam selection map $B^{sorted}$ (line 14).

\begin{figure}[!tp]
	\begin{minipage}{0.2\textwidth}
		\centering
		\includegraphics[width=2.5cm,height=2.5cm]{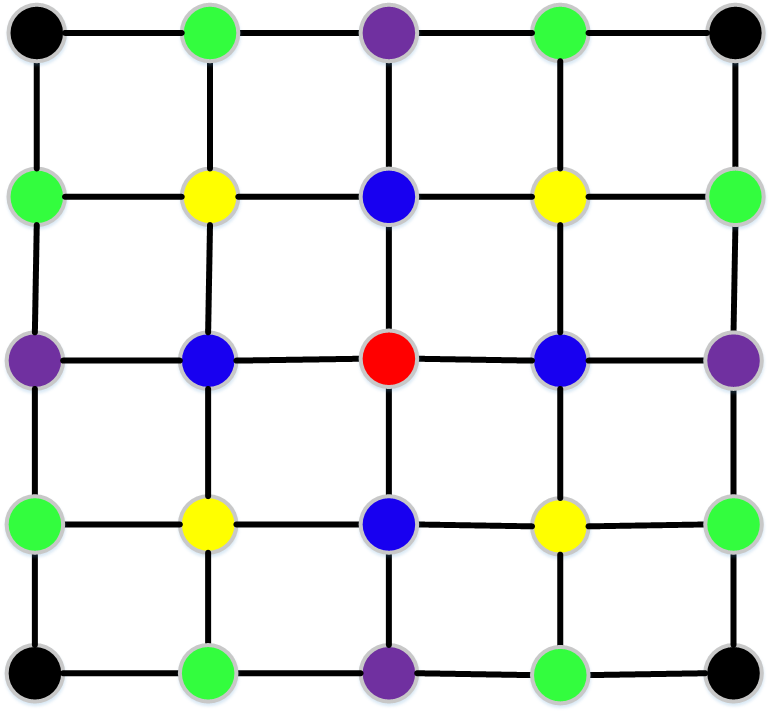}
		\caption{The 2D-grid example}
		\label{fig:2dgrid}
	\end{minipage}
	\begin{minipage}{0.3\textwidth}
		\centering
	    \includegraphics[width=5cm,height=3.5cm]{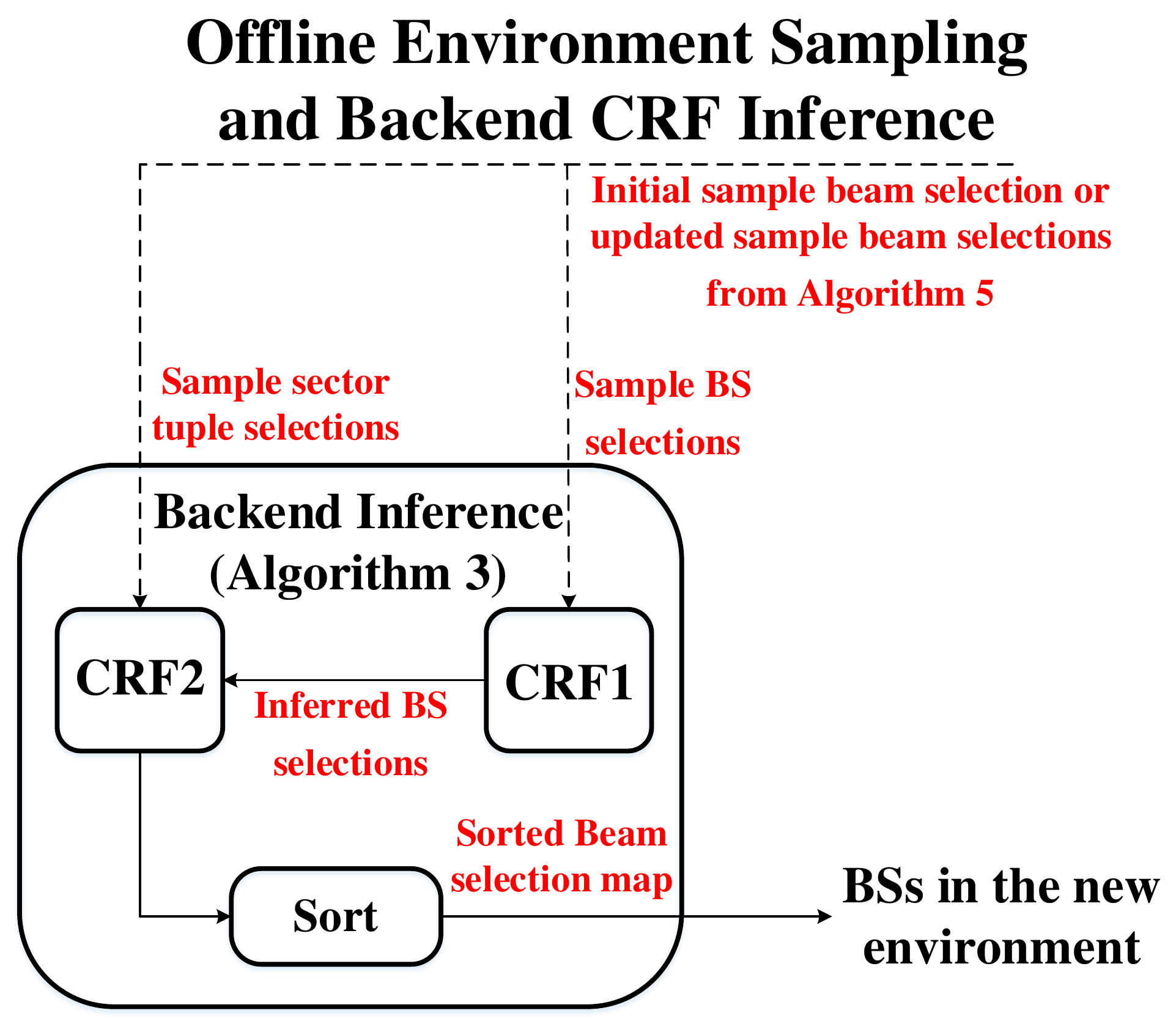}
	    \caption[U-example]{Background CRF Inference}
	    \label{fig:envsampling}
    \end{minipage}
\end{figure}

Fig.~\ref{fig:envsampling} illustrates BIA and environment sampling. For a new environment, sampling devices will be used to find the optimal beam selections at sample locations. Given the trained CRF1 and CRF2, the sampling BS selections will be used as the input of CRF1 and sample sector tuple selections are used as the input to CRF2. CRF1 generates the inferred BS selections in terms of the probability vectors for all the points in the grid (line 3). CRF2 takes the probability vectors and the samples sector tuple selections, generating the conditional probability vectors for every sector tuple selection given every BS selection (line 7). Then the beam selection maps for each points in the grid are derived and sorted (line 13). 

InferBeam uses cascaded CRFs as shown in Fig.~\ref{fig:envsampling}. An alternative design can combine both CRFs. The integrated design can generate a one-shot solution for both BS and sector. However, this integrated design would require to evaluate more possible labels than cascaded design. For the integrated design, the inference procedure would require to evaluate the marginal probability for each of $|X^{BS}|\times |X^{Sec}|$ possible combinations. Whereas for the cascaded design, we just need to evaluate $P(Sec\_ID|BS\_ID)$ only if $P(BS\_ID)$ is high. Moreover, we find from the simulation that for most of the nodes in the grid, $P(BS\_ID)$ does not equal to zeros for only a subset of BSs. Therefore the cascaded design will reduce the overall complexity of inference procedure.
\subsection{Training}
\begin{figure}
	\centerline{\includegraphics[width=7.5cm,height=3cm]{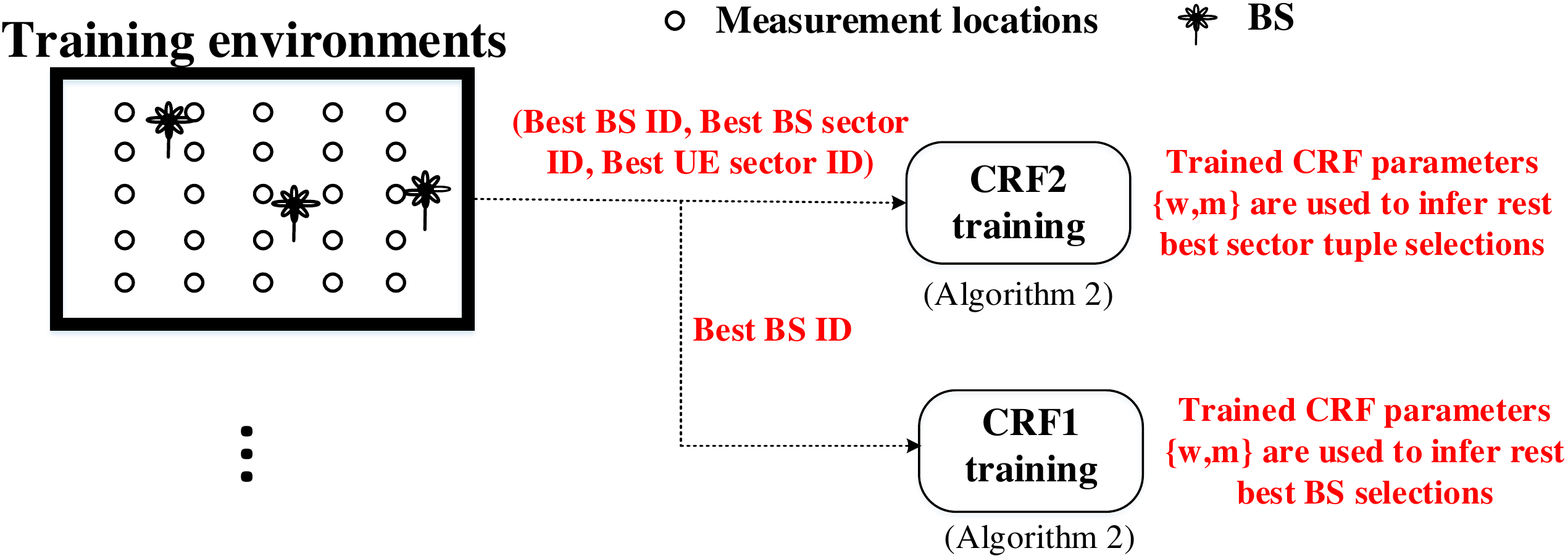}}
	\caption[U-example]{Training procedure for CRF1 and CRF2}
	\label{fig:training}
\end{figure}

Fig.~\ref{fig:training} shows the training procedure for model parameters $\theta={w_{k}, m_{vv'}}$ of the CRF1 and CRF2. The data for training are collected from training environments with a similar dimension as our test environment. Each training environment has the same number of BSs, but they are placed in different locations. In the training data $D = \{D_{r}\}$ and $D^{BS} = \{D^{BS}_{r}\}$, $1\leq r\leq R$, where $R$ is the number of training set, $D_{r}$ and $D^{BS}_{r}$ include the optimal beam selection IDs and optimal BS ID for each small block in the $r$th training environment. CRF1 and CRF2 can be trained with Offline Training Algorithm (OTA) presented in Algorithm 4.

\begin{algorithm}
	\footnotesize
	\textbf{Input}:  $D^{BS}$, $D$, $\theta^{0}$, step size $\eta$, stopping criteria $\Delta$, $I_{max}$ \\
	\textbf{Output}: $\theta^{*} = \{w_{k},m_{vv'}\}$ for CRF1 and CRF2\\
	Use $D^{BS}$ as the input of Algorithm 2 to train CRF1, get the trained parameters $\theta_{1}^{*} = \{w^{*}_{k},m^{*}_{vv'}\},1\leq k\leq K, (v,v')\in E$ for CRF1\\
	Use $D$ as the input of Algorithm 2 to train CRF2, get $\theta_{2}^{*}=\{w^{*}_{k},m^{*}_{vv'}\},1\leq k\leq K, (v,v')\in E$ for CRF2\\
	\textbf{Return} $\theta_{1}^{*}$,$\theta_{2}^{*}$
	\caption{Offline Training Algorithm (OTA)}
\end{algorithm}

OTA uses $D^{BS}$ to train the CRF1 (line 3), and $D$ to train CRF2 (line 4). The algorithm outputs the model parameters for CRF1 and CRF2 (line 5).

\subsection{Overview of BS-UE Beam Alignment Protocol}
We now specify the protocol for beam alignment. First, coarse-grain transmit sectors for BS and UE need to be determined. We choose a set of consecutive transmit sectors for BS and UE as follows. Given a sector range parameter $\xi\in \mathbbm{Z}^+$, we iterate all unique sectors and their coarse-sets by adding in nearby sectors within $\pm \xi$. For example, if $(Sec\_BS\_ID', Sec\_UE\_ID')$ is inferred by CRF2, then the corresponding set of sectors are $([Sec\_BS\_ID'\ominus \xi_{BS}, Sec\_BS\_ID'\oplus \xi_{BS}], [Sec\_UE\_ID'\ominus \xi_{BS}, Sec\_UE\_ID'\oplus \xi_{BS}])$, where $\xi_{BS},\xi_{Sec}$ are the sector rangeS for BS and UE. (Note that modulo addition and subtraction are used because the sectors are placed in circle.) Let $f_{1},f_{2},f_{3},f_{4}$ be $BS\_ID, Sec\_BS\_ID, Sec\_UE\_ID$ and the probability for an entry in $B_v^{sorted}$. The Online Beam Alignment Protocol is presented in Algorithm 5.

\begin{algorithm}	
	\footnotesize
	\textbf{Input}: $B^{sorted}$, $\xi_{BS},\xi_{Sec}$, location estimate of UE, $P_{TH}$\\
	\textbf{Output}: Updated $S'$ and $x^{*}_{S'}$\\
	Set $Update\_with\_erasure = 0$\\
	Find $v'$ in the grid which is closet to the location estimate of UE\\
	\For{each $f \in B^{sorted}_{v'}$}{
		\eIf{$f_{4}>P_{TH}$}{
			$f_{1}$ transmits beacons which contain $f_{3}$ with its sectors $[f_{2}\ominus \xi_{BS},f_{2}\oplus \xi_{BS}]$.\\
			\If{$f_{1}$ receives ACK frame within $RTT$}{
				\textbf{Break}
			}
		}
		{Do SLS with the traditional scheme\\ 	Set $Update\_with\_erasure = 1$ \\ 	\textbf{Break}\\  }		
	}
	Proceed to BRP, find the optimal BS ID and sector ID $(BS\_ID^{*}_{v'}, Sec\_BS\_ID^{*}_{v'}, Sec\_UE\_ID^{*}_{v'})$\\
	\If{$Update\_with\_erasure = 1$}{Append $v'$ to $S$, and $(BS\_ID^{*}_{v'}, Sec\_BS\_ID^{*}_{v'}, Sec\_UE\_ID^{*}_{v'})$ to $x_{S}^{*}$. Clear the samples in $S$ whose physical distance to $v'$ is less than $\beta$.}
	\textbf{Return} $S$ and $x^{*}_{S}$
	\caption{Online BS-UE Beam Alignment Protocol (OBP)}
\end{algorithm}
\begin{figure}
	\centerline{\includegraphics[width=7cm,height=4.5cm]{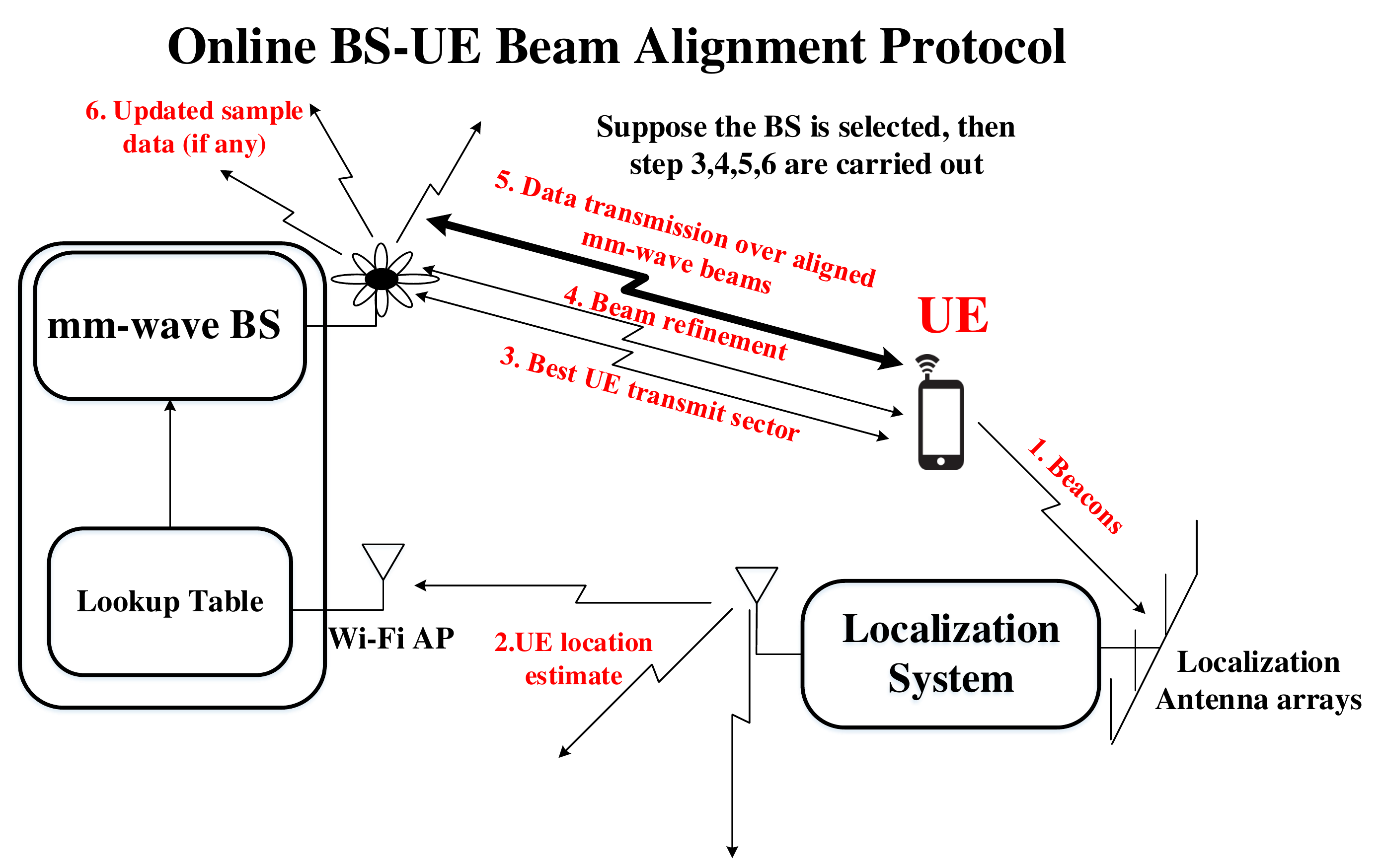}}
	\caption[U-example]{Online BS-UE Beam Alignment Protocol}
	\label{fig:online_protocol}
\end{figure}

A UE first listens in a quasi-omni directional pattern. If all the entries whose probability greater than $P_{TH}$ has been tried (line 6--9), and the connection still does not set up, we use a traditional scheme based on SLS (line 11). If SLS is used, the resulting beam selection will be appended to the sample set $x^{*}_{S}$. The samples with distance to $v'$ less than $\beta$ will be erased (line 16). 

\begin{figure}
	\centerline{\includegraphics[width=9.5cm,height=4.5cm]{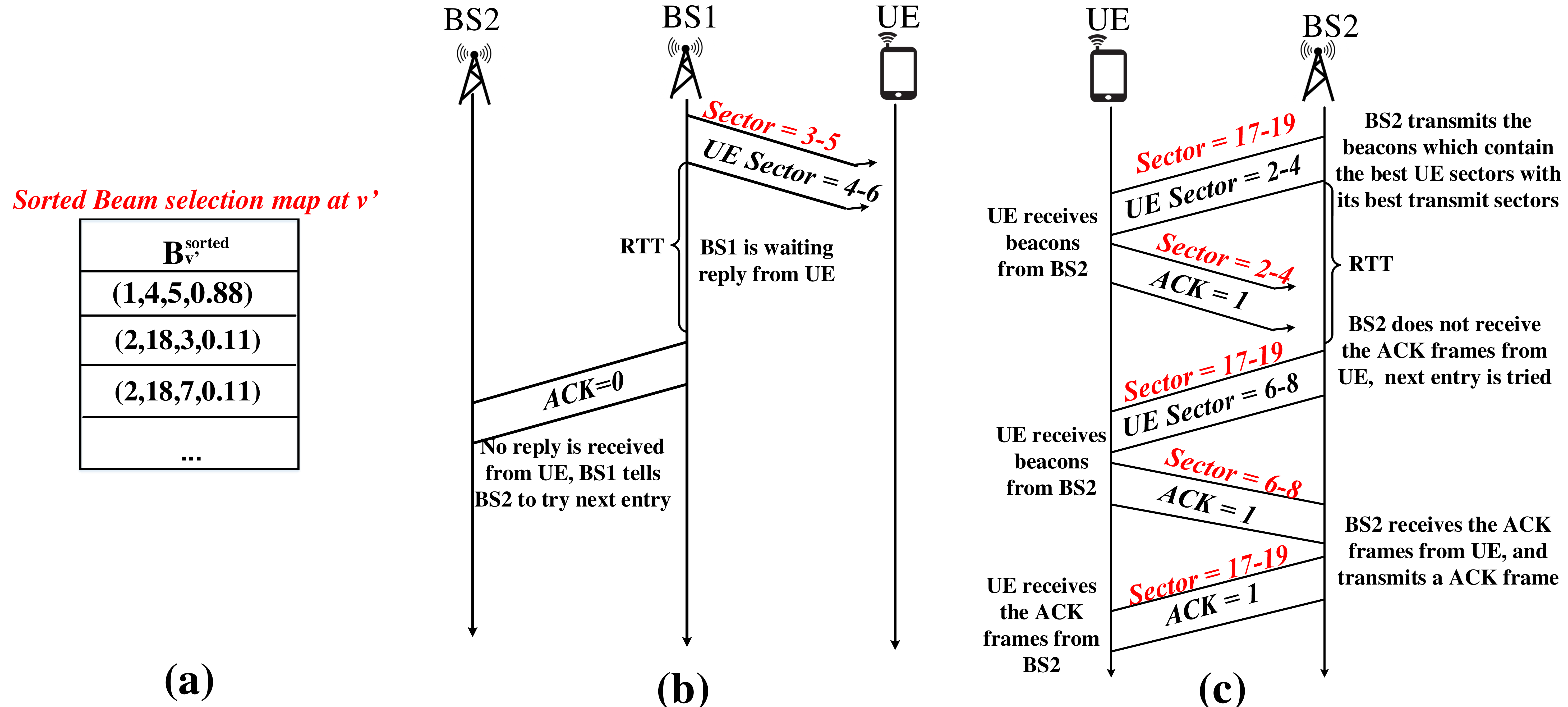}}
	\caption[U-example]{Protocol diagram for sector training}
    \label{fig:online_example}
\end{figure}

Fig. ~\ref{fig:online_example}(a) shows an example of the sorted beam selection map $B^{sorted}_{v'}$. Assume that the mm-wave connections have been set up between every pair of BSs and assume $\xi_{BS} = \xi_{UE} = 1$. In the beginning, UE listens with a quasi-omnidirectional receiver pattern. BS1 transmits the UE transmit sector IDs = 4-6 by using its sectors 3-5 with MCS 0, and it will wait for the reply from UE for a round-trip time (RTT) with a quasi-omnidirectional pattern. Assume UE does not receive the frame. After an $RTT$, BS1 will inform the next best BS (BS2) to connect with UE (Fig. ~\ref{fig:online_protocol}(b)). BS2 will transmit the UE transmit sector IDs = 2-4 with its sectors 17-19 with MCS 0 (Fig. ~\ref{fig:online_protocol}(c)). Assume this time UE receives the frames, UE will send the ACK frame with sectors 17-19, but BS2 does not receive the ACK frame. After $RTT$, BS2 will transmit the UE transmit sector IDs = 6-8 with its sectors 17-19. Assume UE receives the frames and BS2 also receives the ACK frame from UE, BS2 will further send a ACK frame to UE to confirm the connection. 

To infer the beam selection for a new UE, we need the UE location information. For some cases, the localization latency could be an issue. We argue, however, that recent methods \cite{neurallocalization} can achieve a small latency of 0.01\,msec or less, and another method called FILA \cite{FILA} within 10\,msec. Compared with the time needed for connection set up, which can take up to several seconds \cite{linkstateprofile}\cite{outdoormmwave}, the localization latency is insignificant for our purpose.
\subsection{Online Operation for Beam Adjustment}
InferBeam can also dynamically adjust a previously aligned beam when the environment changes (i.e., human blockage). We specify Online Beam Adjustment Protocol (OBAP) in Algorithm 6. 
\begin{algorithm}	
	\footnotesize
	\While{1}{
		Measure the throughput $\alpha_{avg}$ for UE located at $v$ over time period $t$\\
		\If{$\alpha_{avg}<\alpha_{TH}$}{
			UE switches back to Wi-Fi connection mode, and transmits a reconnection signal to all BSs with Wi-Fi\\
			Remove the current entry in use from $B^{sorted}_{v'}$ from the look up table at each BS \\
			Run OBP without remeasure UE's location (line 4), find the next best beam selection
		}
	}
	\caption{Online Beam Adjustment Protocol (OBAP)}
\end{algorithm}

When the average throughput $\alpha_{v}$ of UE drops below a threshold $\alpha_{TH}$, the beam adjustment process is triggered. UE will send a mm-wave reconnection request to all the BSs through Wi-Fi (line 4). The current BS selection, BS transmit sector selection and UE transmit sector selection will be removed from the beam selection map $B_{v'}$ (line 5). OBP then finds the next best beam selection for UE (line 6).

\section{Evaluation}
In this section, we evaluate InferBeam via computer simulation. We use ray-tracing models provided by Wireless InSite \cite{wirelessinsite} to generate our training and test data.
\subsection{Derivation of $\mu_{w_{k}},\mu_{m}$}
\begin{figure}
	\centerline{\includegraphics[width=5.5cm,height=2.5cm]{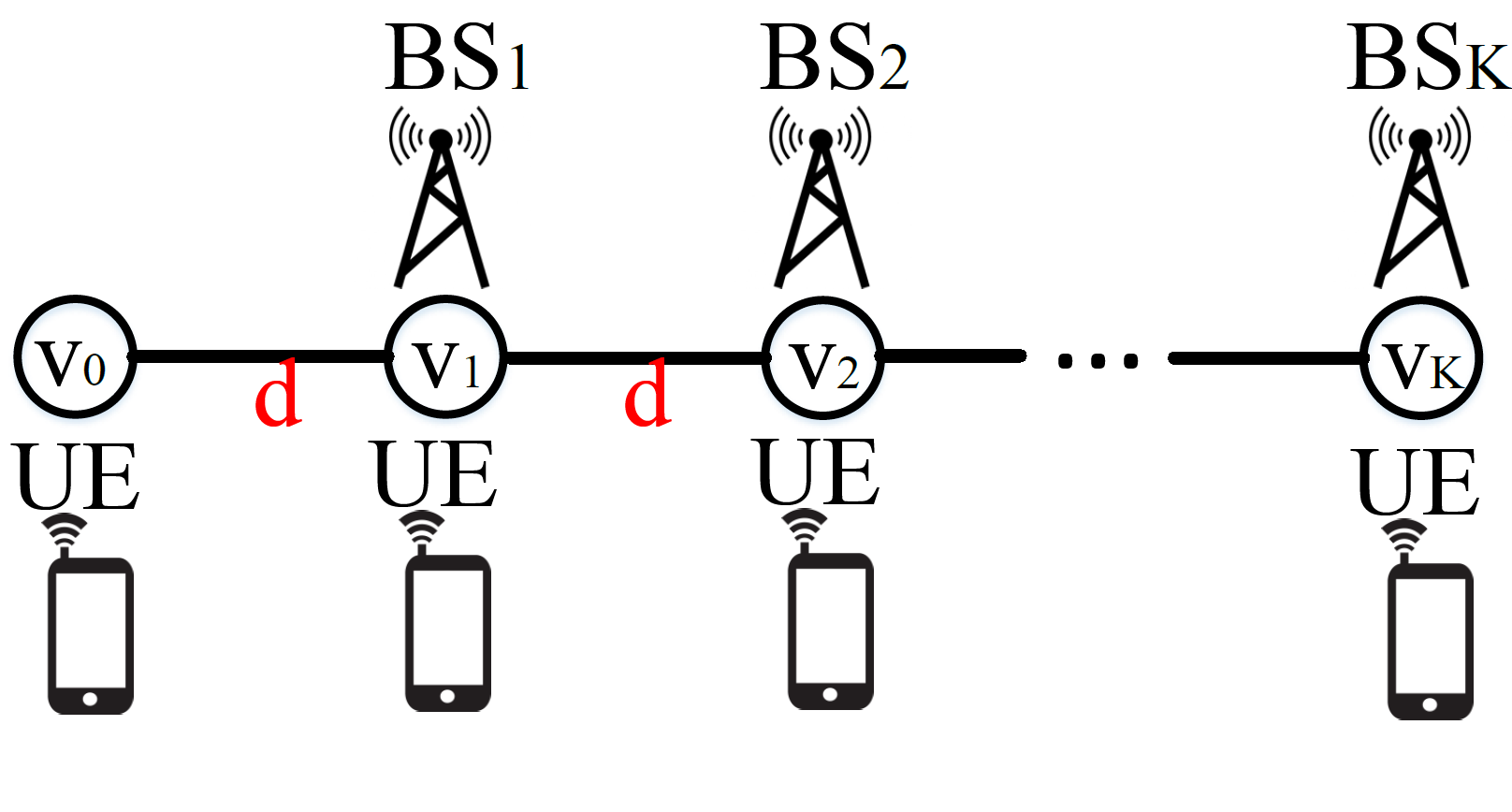}}
	\caption[U-example]{1d grid} 
	\label{fig:1dgrid}
\end{figure}
Considering the simple scenario shown in Fig. \ref{fig:1dgrid}. Assume UEs locate at nodes $v_{0},...,v_{K}$ in the 1D-grid, and mm-wave BSs are placed at $v_{1},...,v_{K}$. Assume each BS transmits beacons and UE listens with an omni-directional antenna pattern. We further assume all the BSs have the same transmitting power and transmitting antenna gain. From \cite{mmmeasure1}, we have the following path loss model for indoor LOS and NLOS scenario: 

\begin{equation}
PL(d)[dB] = PL_{FS}(d_{0})[dB]+10\bar{n}\log_{10}\frac{d}{d_{0}}+X_{\delta}
\end{equation}
where $d_{0} = 1$, $PL_{FS}(d_{0})$ is the free space path loss at $d_{0}=1$, $\bar{n}$ is the \emph{omni-directional path loss exponential} and $X_{\delta}\sim \mathcal{N}(x;0,\delta)$ is the \emph{shadow factor}. $\bar{n}$ and $\delta$ for LOS scenario is 1.1, 1.7 at 28GHz and 1.3,1.9 at 73GHz. For NLOS, $\bar{n}$ and $\delta$ is 2.7, 9.6 at 28GHz and 3.2,11.3 at 73GHz. We take the average of $\bar{n}$ and $\delta$ over two frequencies, getting $\bar{n} = 1.2, \delta=1.8$ for LOS scenario and $\bar{n} = 2.95, \delta=10.45$ for NLOS scenario. 
Assuming that the obstacles are uniformly placed in the environment, $P_{blk}$, the probability that line of sight path between $v_{0}$ and a node is blocked, is proportional to the line of sight distance between them.
\begin{equation}
P_{blk} = P_{e}kd 
\end{equation}
Where $P_{e}$ is the blocking probability per unit distance. $kd$ is the distance between $v_{k}$ and $v_{0}$. Using $(8)$ and neglecting the edge potentials, we have:
\begin{multline}
Q(x_{v_{0}}=BS_{k})\propto exp\Bigg(\sum_{k=0}^{K}w_{k} \smashoperator{\sum_{\substack{\{s|d(s,v_{0})=k\}}}} \mathbbm{1}_{x_{s^{*}}=BS_{k}}\Bigg)\\
=exp(w_{k}\smashoperator{\sum_{\substack{\{s|d(s,v_{0})=k\}}}}\mathbbm{1}_{x_{s^{*}}=BS_{k}}) \times C= exp(w_{k}) \times C
\end{multline}
taking log on both side, we get:
\begin{equation}
w_{k}=\log(Q(x_{v_{0}}=BS_{k}))- log(C)
\end{equation}
where $C$ is a term without $w_{k}$. $(14)$ comes from the fact that $BS_{k}$ is k p-hops away from $v_{0}$. To get $w_{k}$, we need to calculate $Q(x_{v_{0}}=BS_{k})$, the probability that $BS_{k}$ generates the strongest signal at $v_{0}$.
\begin{mydef1}
	The excepted received power (in dB) $E[P_{r}^{k}]$ at $v_{0}$ from $BS_{k}$ is $10\log_{10}(P_{t}G_{t}G_{r})-PL_{FS}(d_{0})-(17.5P_{e}kd+12)\log_{10}(kd)-(X^{NLOS}_{\delta}-X^{LOS}_{\delta_{k}})P_{e}kd-X^{LOS}_{\delta_{k}}$, where $X^{NLOS}_{\delta_{k}}$ and $X^{LOS}_{\delta_{k}}$ are the shadow factors for NLOS and LOS. $P_{t},G_{t},G_{r}$ are the transmitting power, transmitting antenna gain and receiving antenna gain.
\end{mydef1}
\begin{proof}
	See Appendix D.
\end{proof}
Using the result of theorem 4, we can calculate $Q(x_{v_{0}}=BS_{k})$:
\begin{mydef1}
	We have $Q(x_{v_{0}}=BS_{k}) = \prod_{k'=1,k'\neq k}^{K} \mathcal {Q}(\frac{12\log_{10}(\frac{k}{k'})}{1.8\sqrt{2}})$ and $w_{k}=\sum_{k'=1,k'\neq k}^{K}\log(\mathcal{Q}(\frac{12log_{10}(\frac{k}{k'})}{1.8\sqrt{2}})) + C$. Where $\mathcal {Q}(.)$ is the Q-function of Gaussian random variable and $C$ is the term which does not depend on $w_{k}$. 
\end{mydef1}
\begin{proof}
	See Appendix E.
\end{proof}
Because we are only interested in the ratio between $exp(w_{k})$, we set $C=0$. This represents our prior belief in the value of $w_{k}$, therefore we make $\mu_{w_{k}} = \sum_{k'=1,k'\neq k}^{K}\log(\mathcal{Q}(\frac{12\log_{10}(\frac{k}{k'})}{1.8\sqrt{2}}))$.

To calculate $\mu_{m}$, we have the following derivation: $\frac{Q(x_{v_{0}}=BS_{2})}{Q(x_{v_{0}}=BS_{1})} = \frac{Q(x_{v_{0}}=BS_{2})\prod_{v\in V,v\neq v_{0}}Q(x_{v})}{Q(x_{v_{0}}=BS_{1})\prod_{v\in V,v\neq v_{0}}Q(x_{v})} \approx \frac{P(x_{v_{0}}=BS_{2}, x_{V\backslash \{v_{0}\}})}{P(x_{v_{0}}=BS_{1}, x_{V\backslash \{v_{0}\}})}$.  
Neglect the difference on $\phi(x_{v_{0}} = BS_{1})$ and $\phi(x_{v_{0}} = BS_{2})$, and assume $m_{ij} =\mu_{m}$, $\forall (i,j)\in E$, the above equation equals $\frac{\psi(x_{v_{0}} = BS_{2},x_{v_{1}} = BS_{1})}{\psi(x_{v_{0}} = BS_{1},x_{v_{1}} = BS_{1})} = exp(\mu_{m})$. Therefore we have $\frac{Q(x_{v_{0}}=BS_{2})}{Q(x_{v_{0}}=BS_{1})} \approx exp(\mu_{m})$ and $\mu_{m} = \log(\frac{Q(x_{v_{0}}=BS_{2})}{Q(x_{v_{0}}=BS_{1})})$. Substituting the expression for $Q(x_{v_{0}}=BS_{1})$ and $Q(x_{v_{0}}=BS_{2})$, we have $\mu_{m} = \frac{\sum_{k'=1,k'\neq 2}^{K} \log \mathcal{Q}(\frac{12\log_{10}(\frac{2}{k'})}{1.8\sqrt{2}})}{\sum_{k'=1,k'\neq 1}^{K} \log \mathcal{Q}(\frac{12\log_{10}(\frac{1}{k'})}{1.8\sqrt{2}})}$. 
\subsection{Experimental methodology}

\begin{figure}
	\centerline{\includegraphics[width=8.5cm,height=1.2cm]{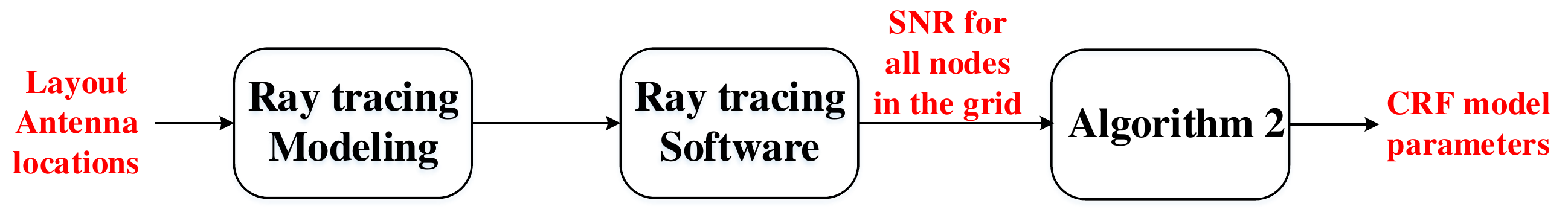}}
	\caption[U-example]{Training procedure}
	\label{fig:blockdiagram}
\end{figure}
\begin{figure}
	\centerline{\includegraphics[width=6.5cm,height=2.5cm]{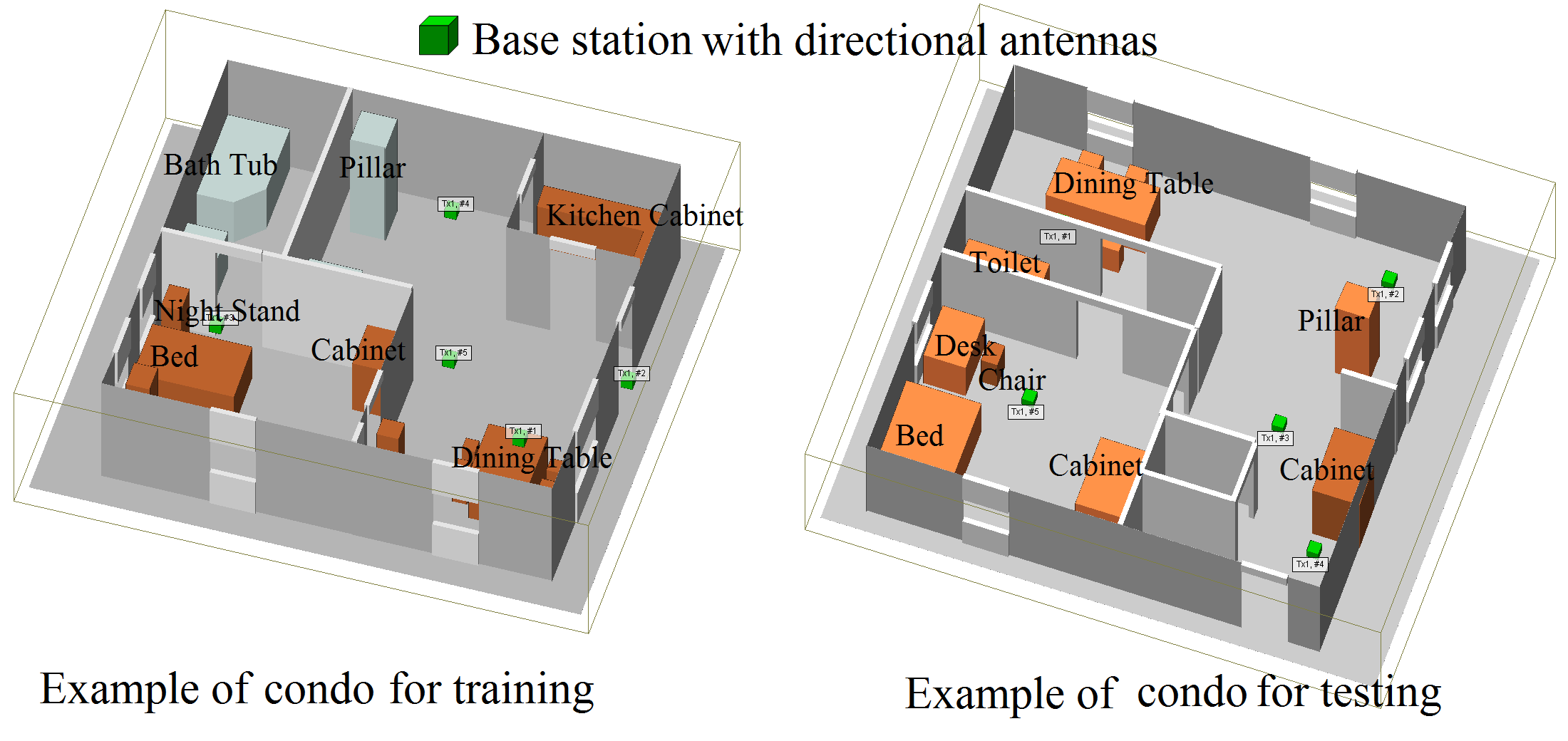}}
	\caption[U-example]{Examples of layouts of condo}
	\label{fig:condolayout}
\end{figure}
\begin{figure}
	\centerline{\includegraphics[width=7.2cm,height=2.5cm]{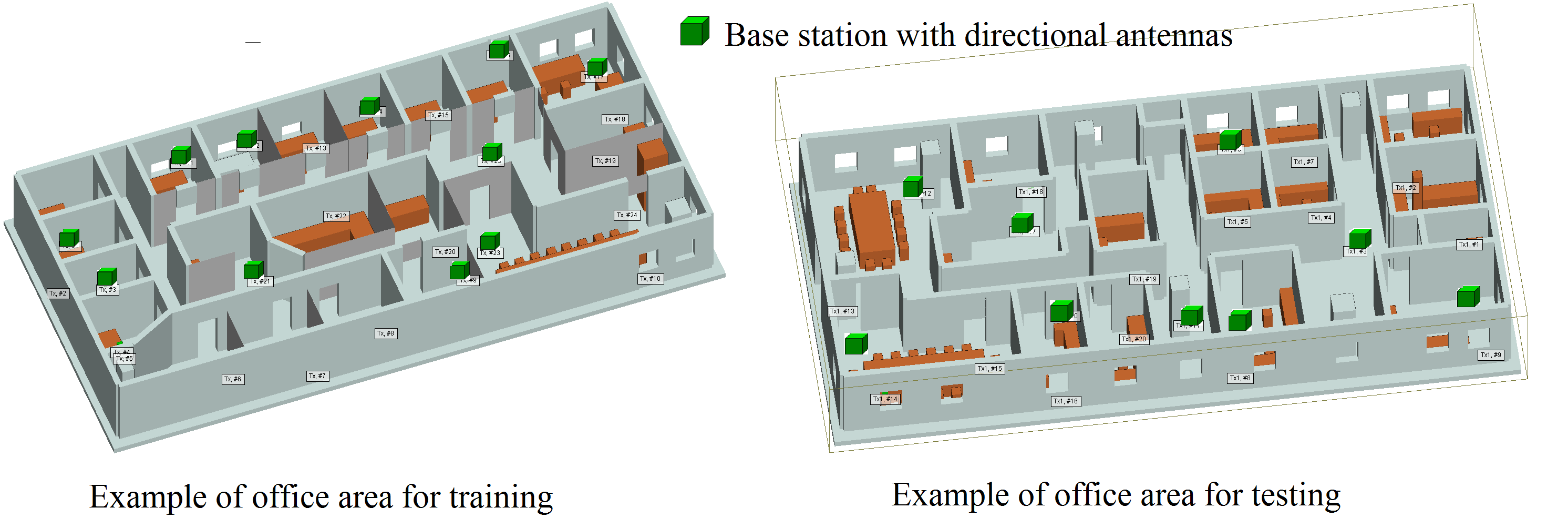}}
	\caption[U-example]{Examples of layouts of office area}
	\label{fig:officelayout}
\end{figure}

Training with simulated data is illustrated in Fig.~\ref{fig:blockdiagram}. The evaluation data are generated from two different types of environments, condominium and office, to simulate typical living and working spaces. We use 55 floor plans for each type from \cite{onebedroomfloorplan} and \cite{officeareafloorplan}. All of the condos have an approximate dimension of $8.5m\times 8m\times 3m$, and the offices have $70m\times 40m\times 4m$. We use 50 out of all 55 floor plans for training and 5 for testing. Examples of the layouts for training and testing are shown in Figs.~\ref{fig:condolayout} and \ref{fig:officelayout}.

We overlay the 3D grid over all environments using a block of $15cm\times 15cm\times 15cm$. We place 5 BSs at random locations for each condo. To measure the signal strength at each location, we place a UE at the center of each small block. Each BS antenna and UE antenna has 60 transmit sectors. For offices, because of its larger size, we use a larger block of $40cm\times 40cm\times 40cm$ for the 3D grid. The detailed simulation settings are summarized in Table I. 


To find the optimal transmit sector ID from BS to UE, each BS transmits beacons at 60 GHz from its sectors while UE listens with a omni-directional antenna pattern. The IDs of the BS and the ID of its transmit sector that generates the highest SNR at UE is recorded. The procedure is repeated similarly to find the best UE transmit sector. After getting the optimal beam selection for each of the nodes in the grid, we train CRF1 and CRF2 using Algorithm 4.

We make the prior distribution of $w_{k}$ and $m_{vv'}$ to be Gaussian. We use a simple example to find an analytic solution for $\mu_{w_{k}}$ and $\mu_{m}$. We get $mu_{w_{k}}= \sum_{k'=1,k'\neq k}^{K}\log(\mathcal{Q}(\frac{12\log_{10}(\frac{k}{k'})}{1.8\sqrt{2}}))$ and $\mu_{m} = \frac{\sum_{k'=1,k'\neq 2}^{K} \log \mathcal{Q}(\frac{12\log_{10}(\frac{2}{k'})}{1.8\sqrt{2}})}{\sum_{k'=1,k'\neq 1}^{K} \log \mathcal{Q}(\frac{12\log_{10}(\frac{1}{k'})}{1.8\sqrt{2}})}$. The derivation can be found in \cite{tech_report}.
\begin{table}[tp!]
	\scriptsize
	\caption{Simulation Settings} 
	\centering 
	\begin{tabular}{p{2.8cm} p{1.4cm} p{1.6cm}}
		\hline 
		Name & Condominium& Office area\\ 
		\hline 
		Number of training data & 50 & 50 \\
		Number of test data & 5  &  5 \\ 
		Number of BSs& 5 & 20 \\ 
		Number of sectors per BS& 60 &60 \\
		Number of sectors per UE& 60 & 60 \\
		Dimension $(m^{3})$ & $8.5 \times 8\times 3$  & $70 \times 40\times 4$ \\
		Number of nodes in the grid & $56\times53 \times 20$ & $175 \times100 \times 10$ \\
		Number of samples & $350$ & $600$ \\
		\hline 
	\end{tabular}
	\label{table:nonlin} 
\end{table}

\subsection{Baseline results}
\begin{figure}[!tp]
	\begin{minipage}{0.15\textwidth}
		\centering
		\includegraphics[width=2.9cm,height=2cm]{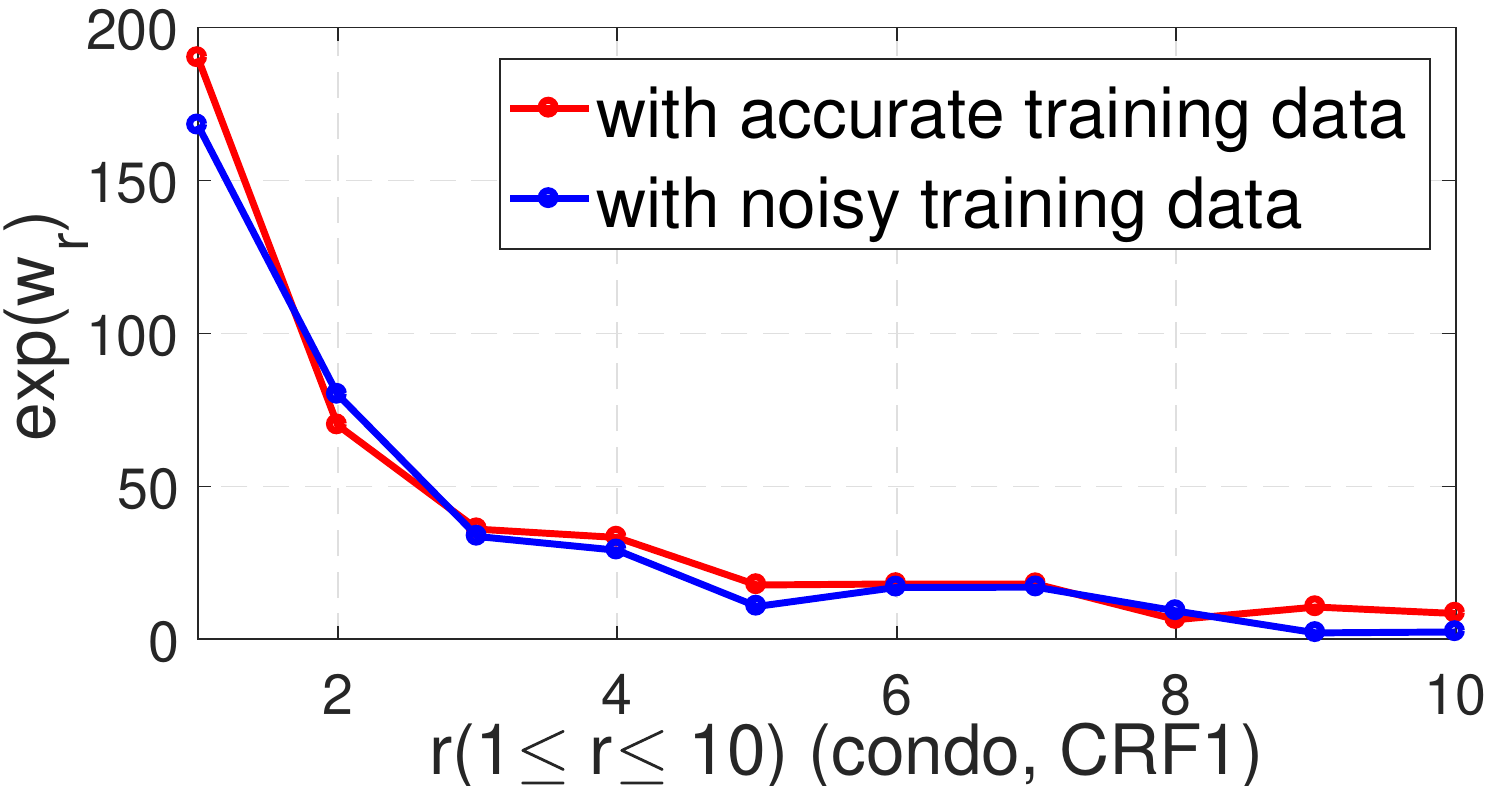}
		\caption{$w_{r}$ of CRF1 for condo}
		\label{fig:wcrf1condo}
	\end{minipage}
	\begin{minipage}{0.15\textwidth}
		\centering
		\includegraphics[width=2.9cm,height=2cm]{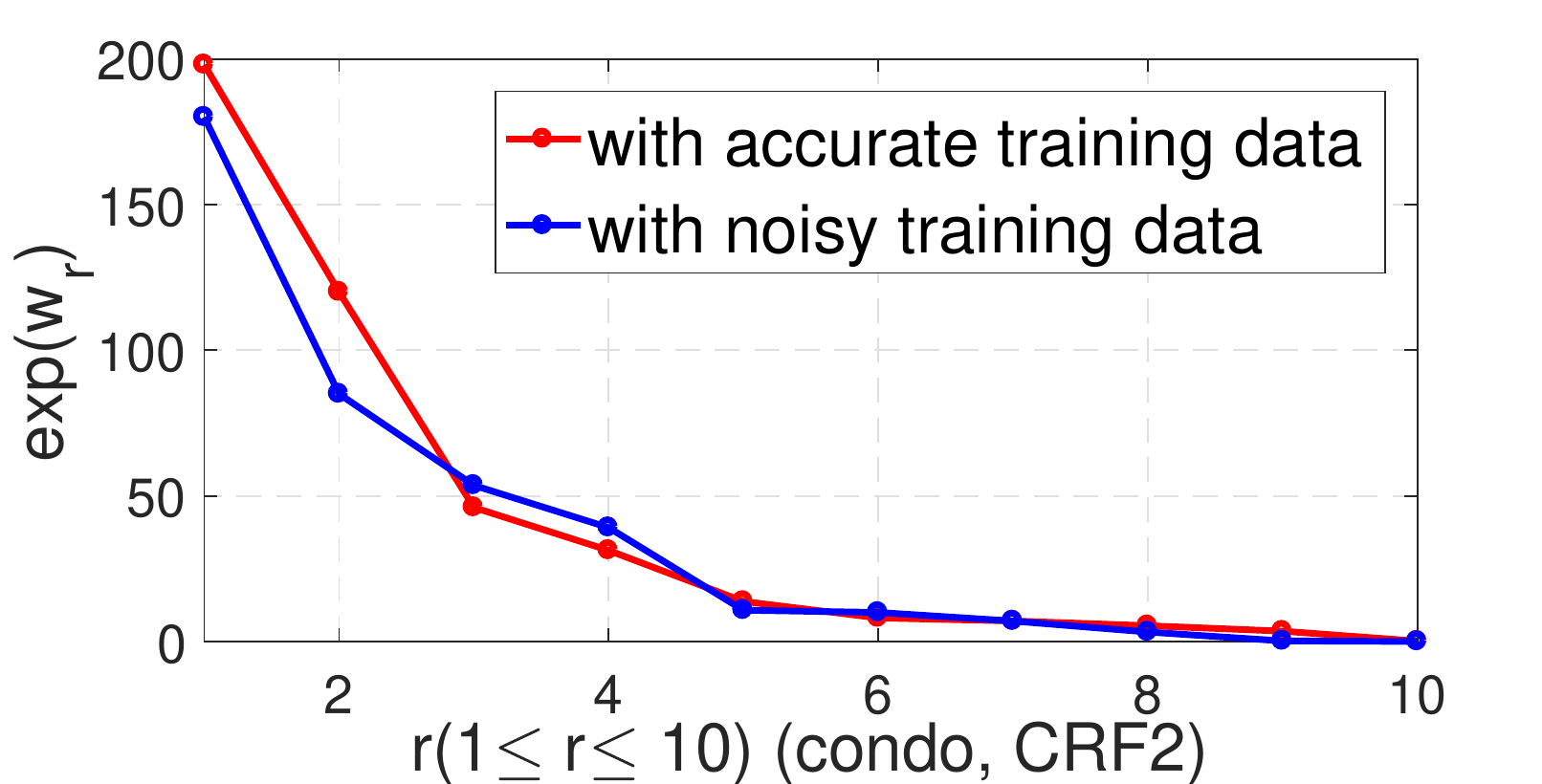}
		\caption{$w_{r}$ of CRF2 for condo}
		\label{fig:wcrf2condo}
	\end{minipage}
	\begin{minipage}{0.15\textwidth}
		\centering
		\includegraphics[width=3.1cm,height=2cm]{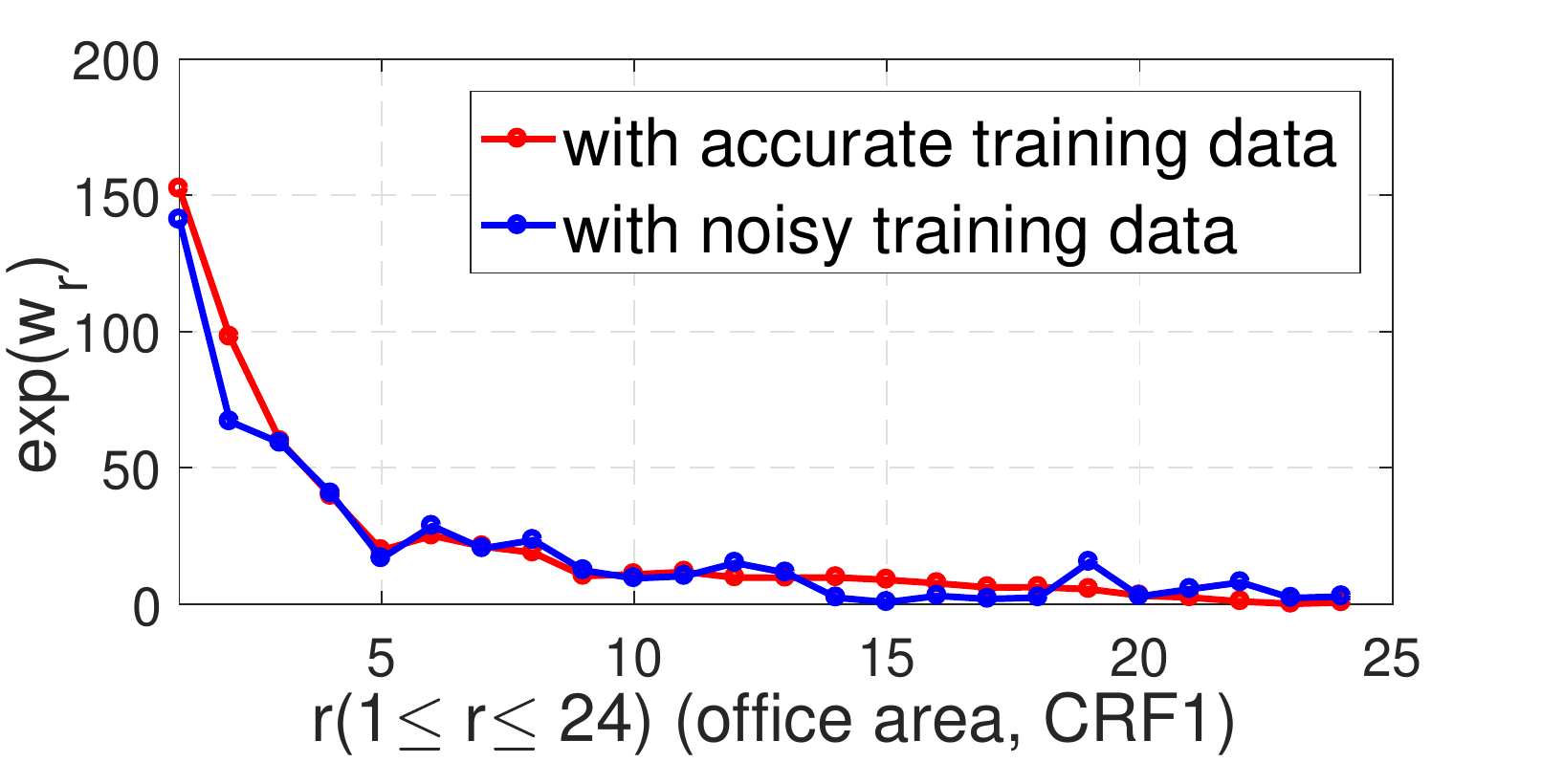}
		\caption{$w_{r}$ of CRF1 for office area}
		\label{fig:wcrf1office}
	\end{minipage}
\end{figure}

\begin{figure}[!tp]
    \begin{minipage}{0.15\textwidth}
			\centering
			\includegraphics[width=3cm,height=2cm]{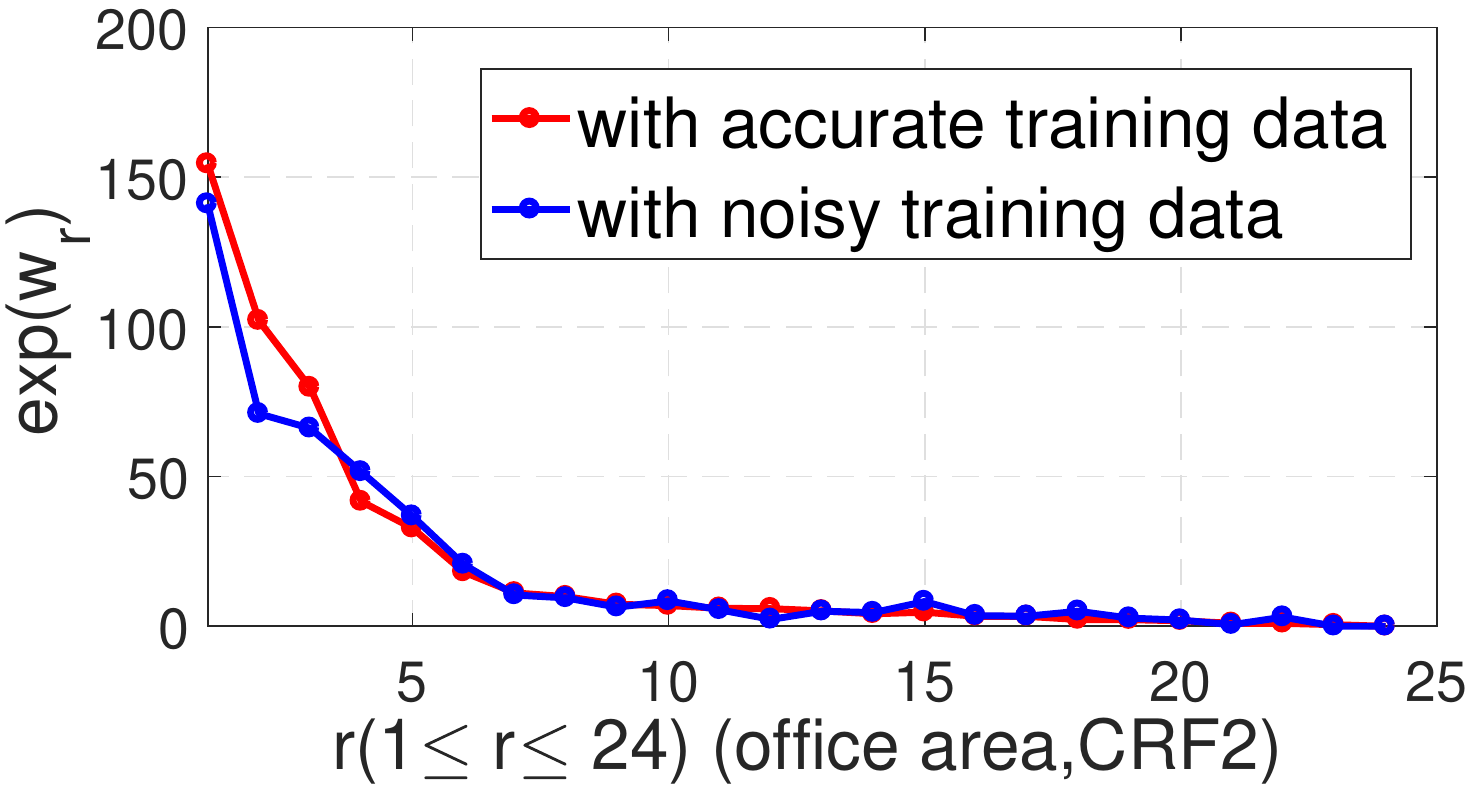}
			\caption{$w_{r}$ of CRF2 for office area}
			\label{fig:wcrf2office}
	\end{minipage}
	\begin{minipage}{0.15\textwidth}
		\centering
		\includegraphics[width=2.9cm,height=2cm]{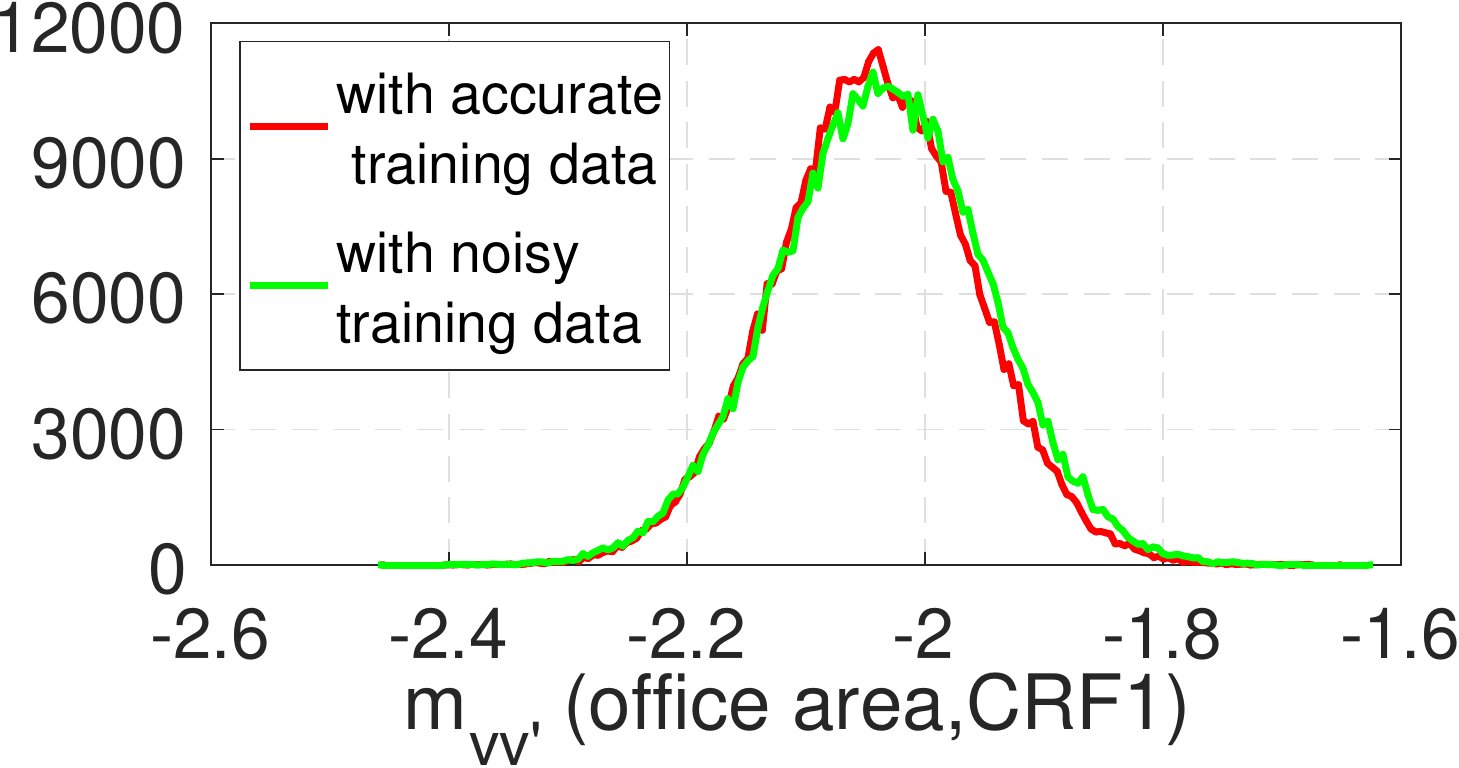}
		\caption{$m_{vv'}$ for office area (CRF1)}
		\label{fig:mcrf1office}
	\end{minipage}
	\begin{minipage}{0.15\textwidth}
		\centering
		\includegraphics[width=3cm,height=2.2cm]{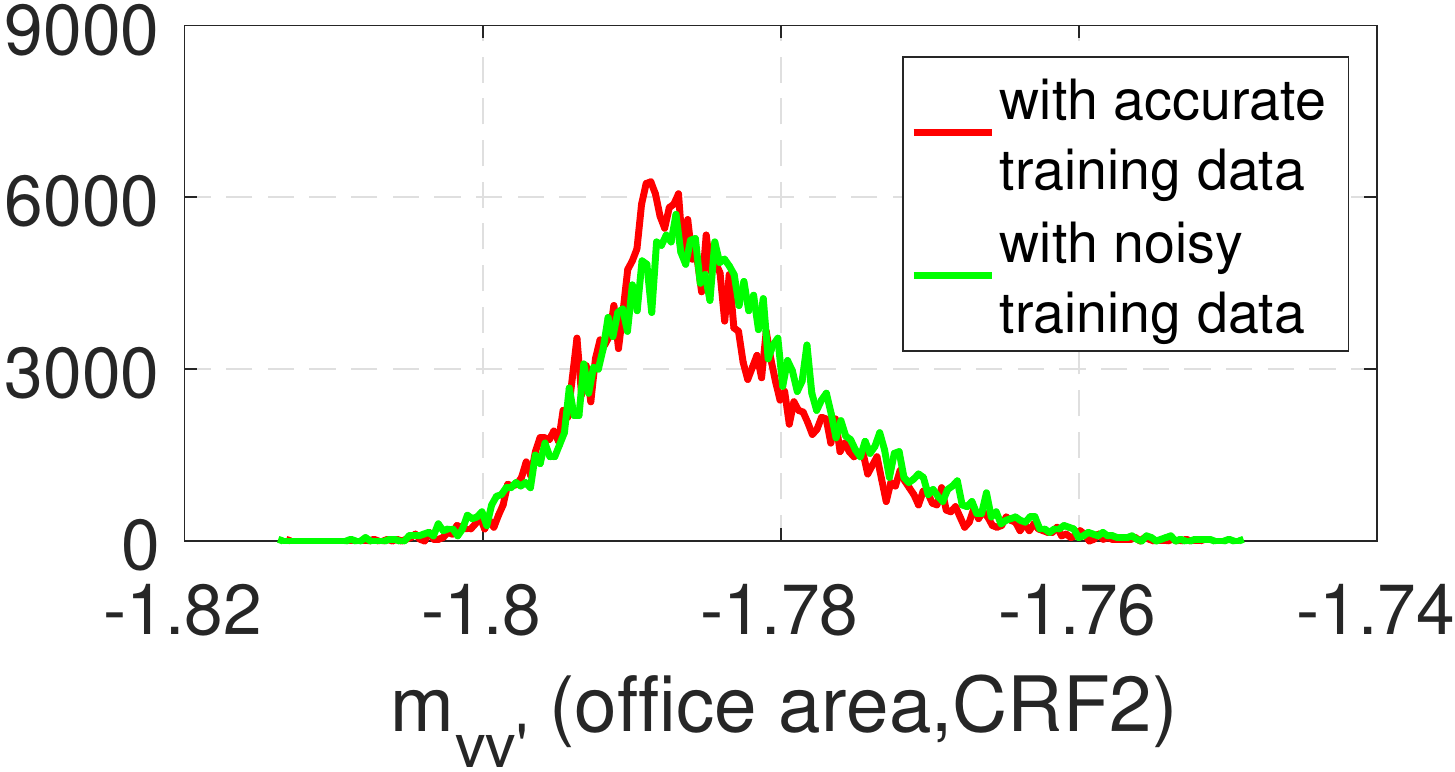}
		\caption{$m_{vv'}$ for office area (CRF2)}
		\label{fig:mcrf2office}
	\end{minipage}
\end{figure}

Figs. \ref{fig:wcrf1condo}--\ref{fig:wcrf2office} plot only the $exp(w_{r})$ part of the node potential functions that are larger than 0.1. We neglect the rest of $w_{r}$ because they do not affect the inference outcome. Here, we examine $w_{r}$ trained on clean (i.e., noiseless) data. First, we notice that $w_{r}$ decays faster for condo than office area. This is due to the effect that BS and the obstacles (furnitures) in the condo are distributed more densely than office area, hence abrupt changes on received signal strengths at the grid nodes are more frequent. This will cause $w_{r}$ which is a measure of spatial correlation on ground truth beam selection to be smaller. Second, $w_{r}$ of CRF2 decays faster than that of $CRF1$ for both environments. The reason is the number of ground truth labels (BS for CRF1, and sector tuples for CRF2) is larger for CRF2 than CRF1, the changes on ground truth label of the grid points is more frequent for CRF2 than CRF1, which causes the $w_{r}$ of CRF2 smaller.

Figs. \ref{fig:mcrf1office}-\ref{fig:mcrf2condo} plot the histograms of the $m_{vv'}$ of CRF1 and CRF2 for both environments. We examine $m_{vv'}$ trained on clean data. We notice $m_{vv'}$ is greater in general for CRF2 than CRF1. The reason is that the number of ground truth labels of CRF2 is larger, and the changes on ground truth label of the grid points is more frequent for CRF2 than CRF1, which causes the $m_{vv'}$, a measure of label inconsistency between the consecutive nodes, larger for CRF2.

\subsection{Results with noisy training data}
\begin{figure}[!tp]
	\begin{minipage}{0.15\textwidth}
		\centering
		\includegraphics[width=2.9cm,height=2cm]{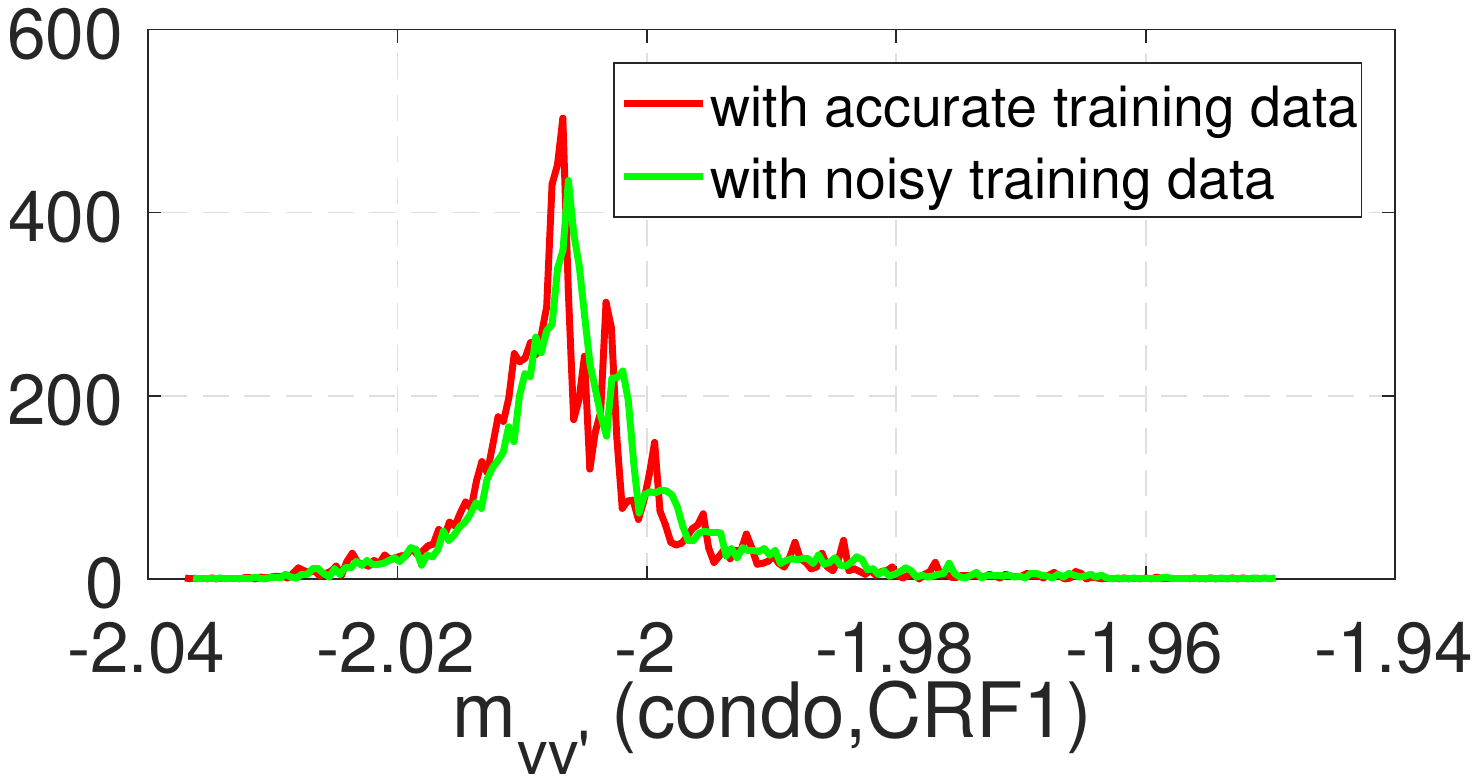}
		\caption{$m_{vv'}$ of CRF1 for condo}
		\label{fig:mcrf1condo}
	\end{minipage}
	\begin{minipage}{0.15\textwidth}
		\centering
		\includegraphics[width=3cm,height=2cm]{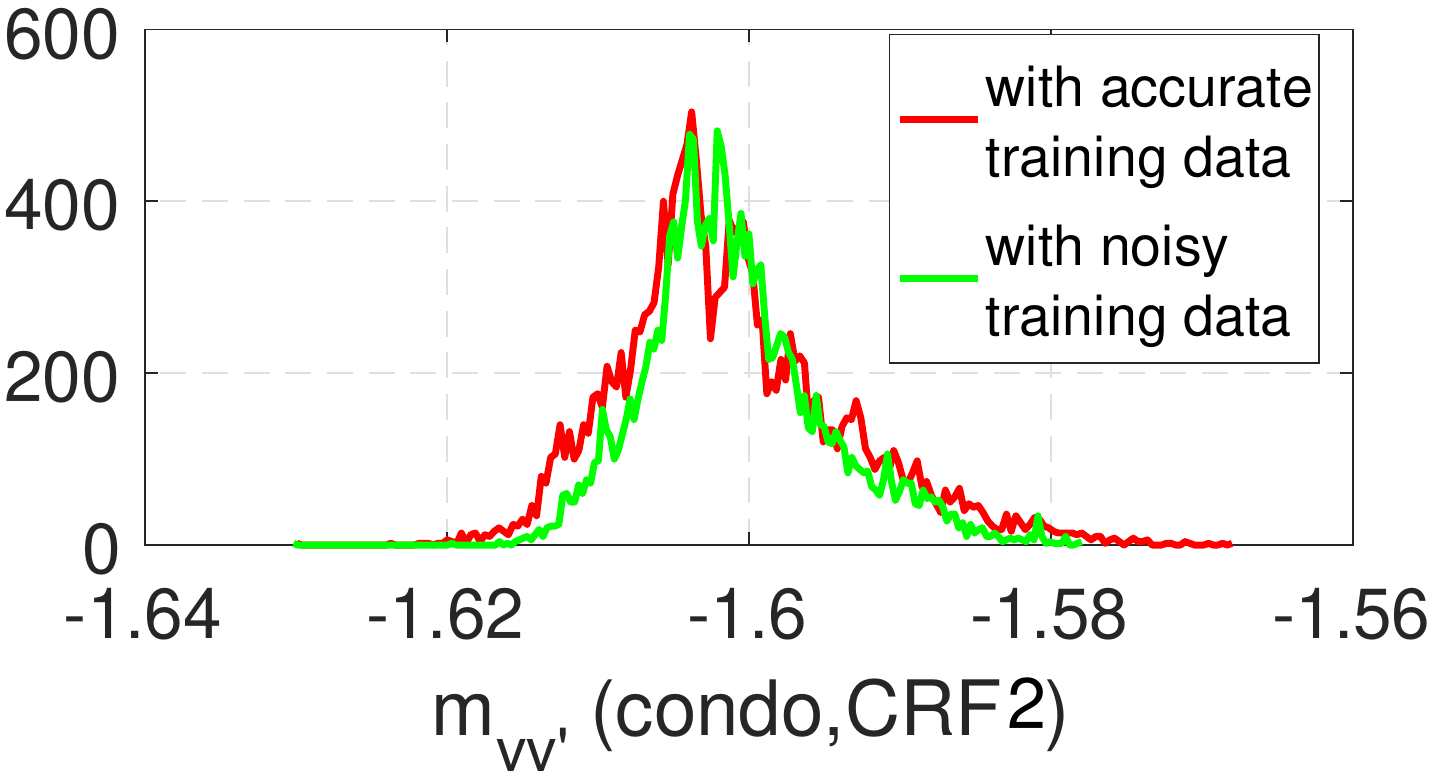}
		\caption{$m_{vv'}$ of CRF2 for condo}
		\label{fig:mcrf2condo}
	\end{minipage}
	\begin{minipage}{0.15\textwidth}
		\centering
		\includegraphics[width=3cm,height=1.9cm]{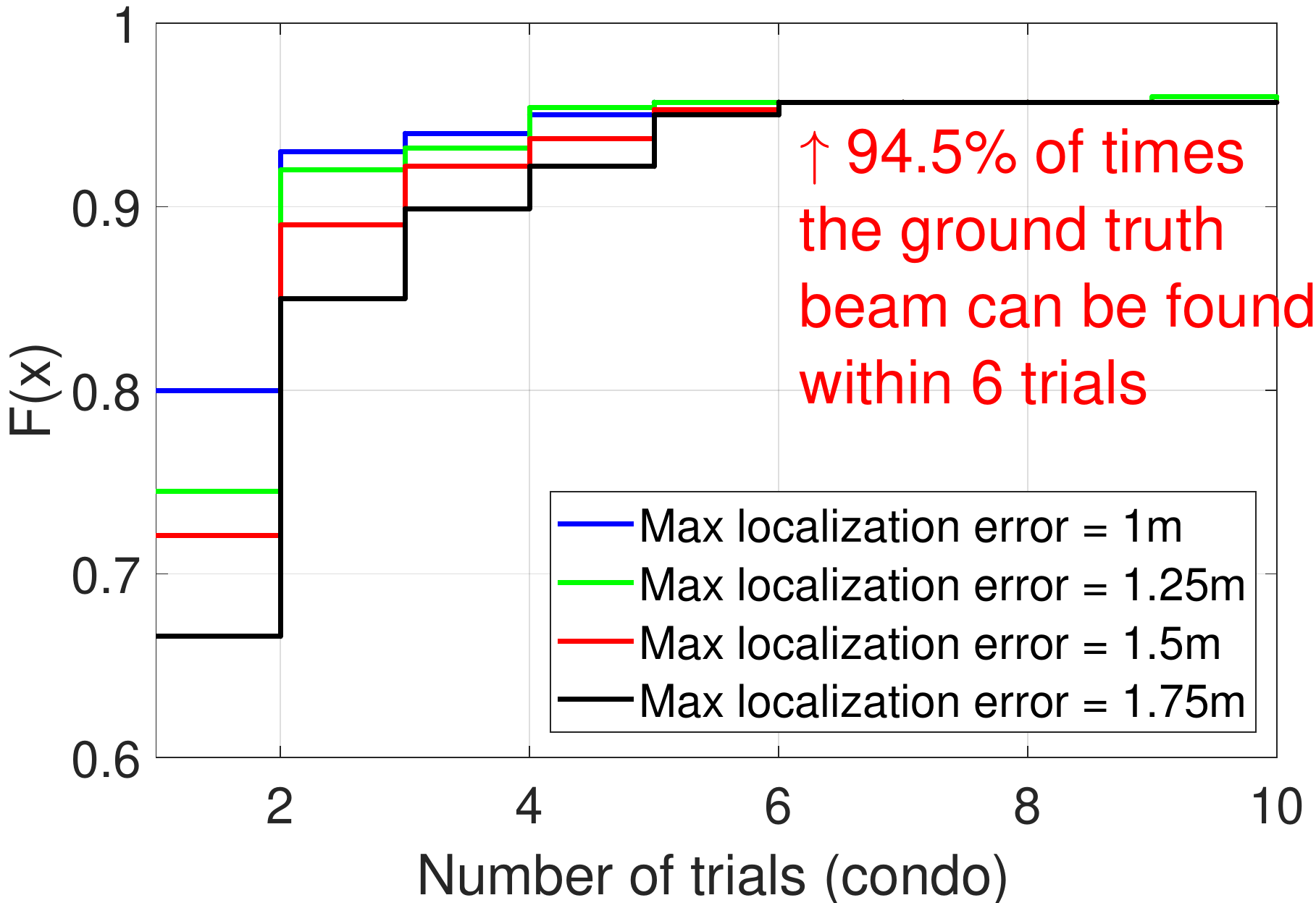}
		\caption{CDF of trials with different localization accuracy for condo}
		\label{fig:cdf1}
	\end{minipage}
\end{figure}

We now examine the effect of noisy training data. We add Gaussian noise to the received clean signals at UE and BS at every training set such that $17\%$ of the nodes in the grid change their ground truth $BS\_ID$, and $20\%$ of the nodes in the grid change their ground truth $Sec\_ID$ on average. Then, CRF1 and CRF2 are trained with the noisy training data using Algorithm 4. Figs. \ref{fig:wcrf1condo}--\ref{fig:wcrf2office} show the values of $exp(w_{r})$ of CRF1 and CRF2 trained with clean and noisy data for comparison. We notice the difference insignificant. A reason is that the gradient $\frac{\partial \frac{1}{R}log P(\theta|D)}{\partial w_{k}}$ used by Algorithm 2 to find the optimal value of $w_{k}$ only depends on $\sum_{\{s|d(s,v)=k\}} \mathbbm{1}_{x^{*}_{s}=x_{v}}$, which is the total number of matches on $x_{s*}$ and $x_{v}$ for the $s$ which are $k$ p-hops from $v$. Even if the noise changes $x_{s*}$ and $x_{v}$, $\sum_{\{s|d(s,v)=k\}} \mathbbm{1}_{x^{*}_{s}=x_{v}}$ may keep the same or change less.
Figs. \ref{fig:mcrf1office}-\ref{fig:mcrf2condo} show the histograms of $m_{vv'}, (v,v')\in E$ for CRF1 and CRF2 in different environments. The difference is also not too significant, the reason is similar, gradient $\frac{\partial \frac{1}{R}log P(\theta|D)}{\partial m_{vv'}}$ used by Algorithm 2 to find the optimal value of $m_{vv'}$ only depends on $\mathbbm{1}_{x_{v}\neq x_{v'}}$, not the exact value of $x_{v}$ and $x_{v'}$. Therefore, we conclude that the learned parameters are insensitive to noise in the
training data. Thus, we can expect that CRF model is highly transferable to different in-door environments. This is demonstrated in our test scenarios where the test condos can have different layouts from training ones.
\subsection{Effect of localization errors}
\begin{figure}[!tp]
	\begin{minipage}{0.15\textwidth}
		\centering
		\includegraphics[width=2.8cm,height=2cm]{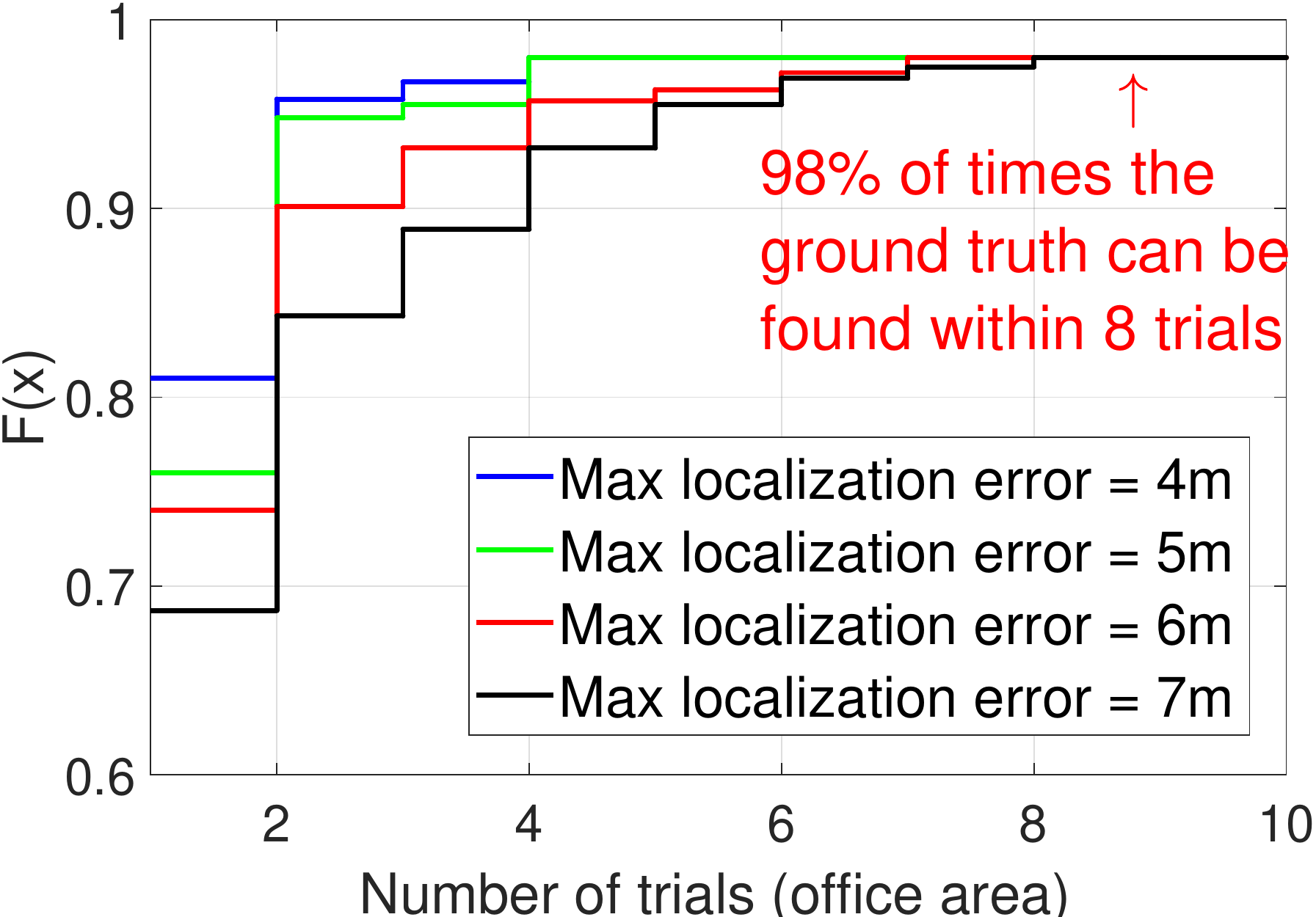}
		\caption{CDF of trials with different localization accuracy for office}
		\label{fig:cdf2}
	\end{minipage}
	\begin{minipage}{0.15\textwidth}
		\centerline{\includegraphics[width=2.8cm,height=2cm]{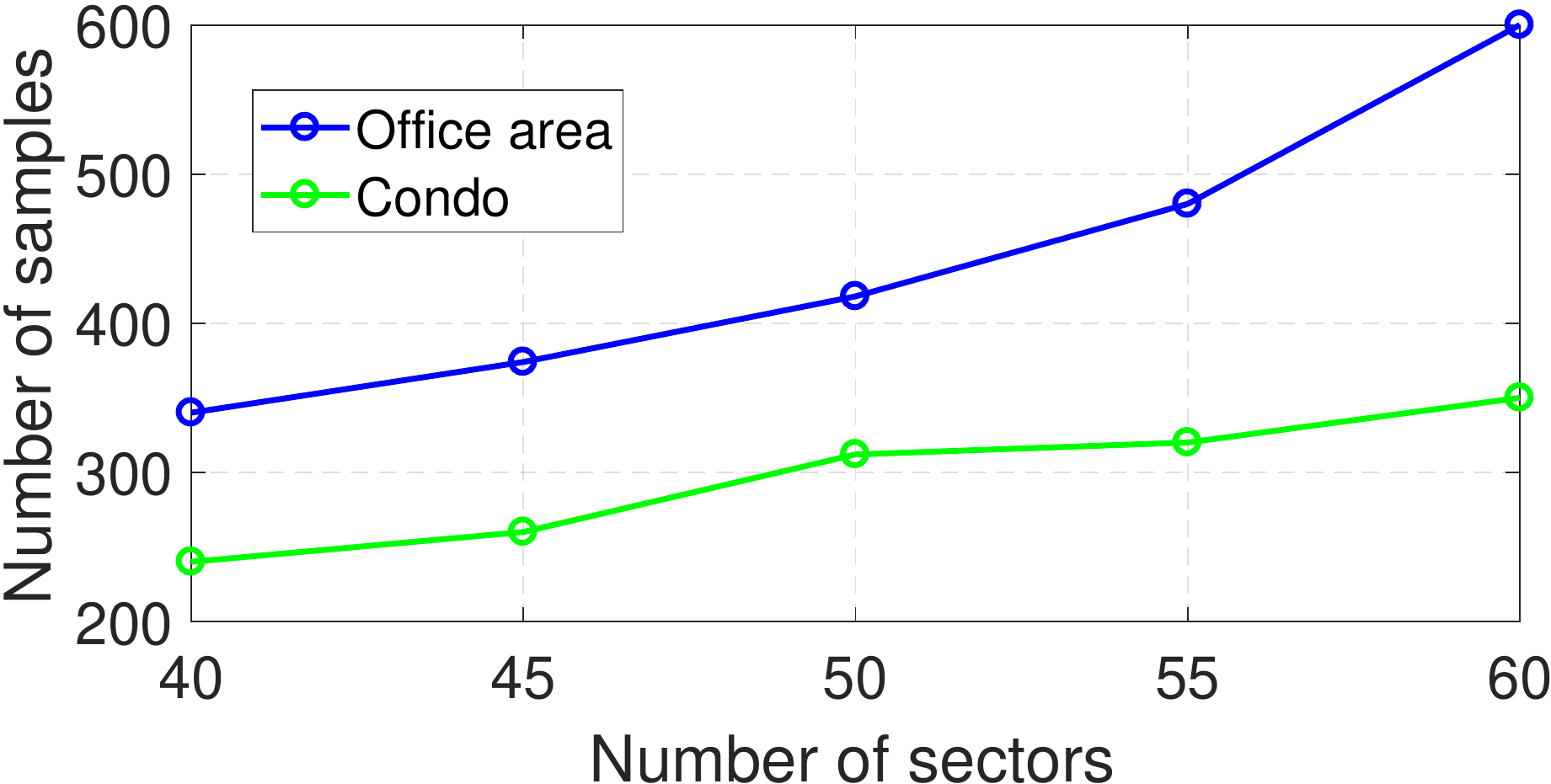}}
		\caption[U-example]{Number of samples needed for different number of sectors}
		\label{fig:numbersamples}
	\end{minipage}
	\begin{minipage}{0.15\textwidth}
		\centerline{\includegraphics[width=2.9cm,height=2cm]{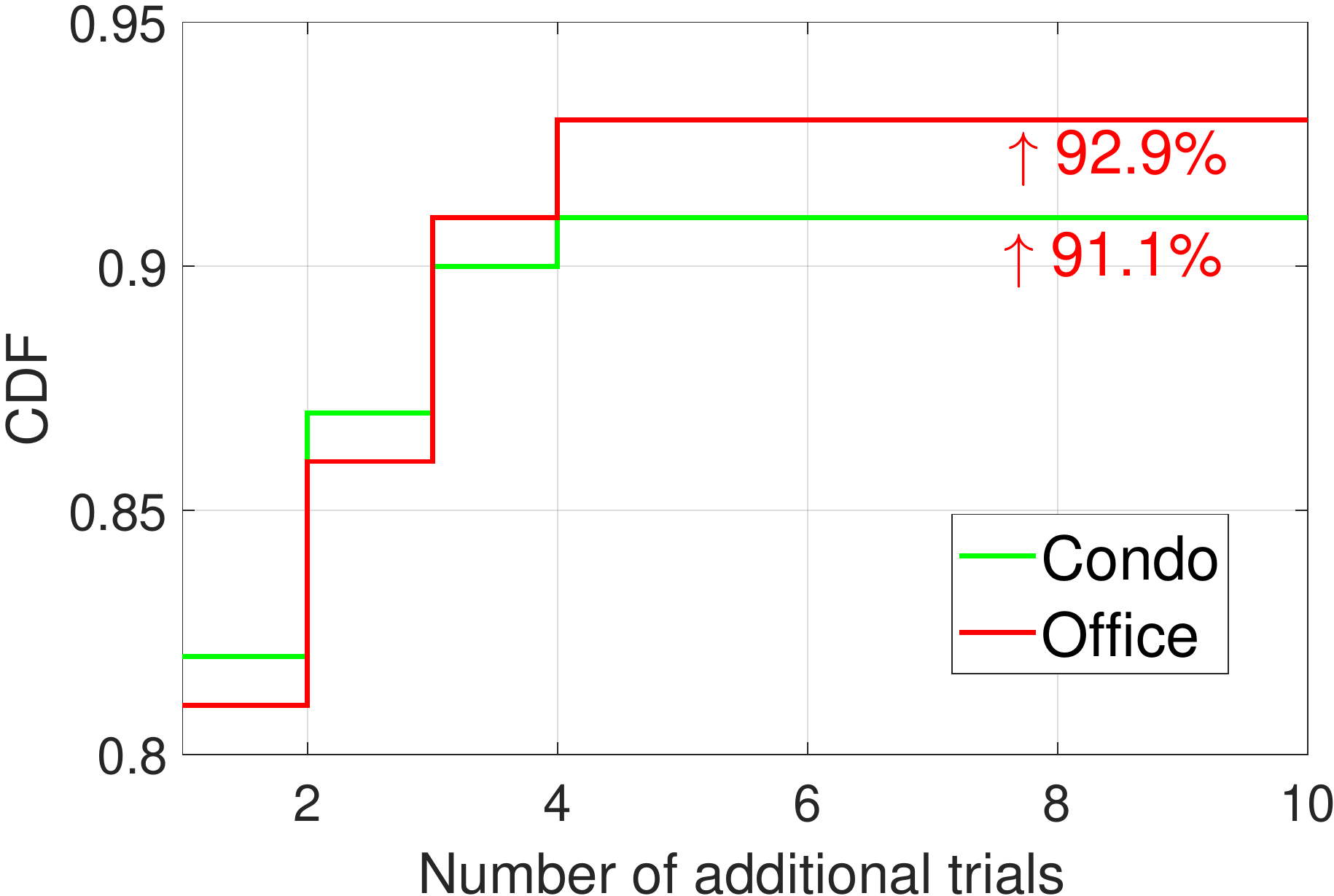}}
		\caption[U-example]{Number of additional trials for beam adjustment}
		\label{fig:additionalsamples}
	\end{minipage}
\end{figure}

We analyze the accuracy of beam inference returned by OBP with different localization errors. The current indoor localization systems has a localization error less than 1m for a small indoor area (condo)\cite{smartcondo} and 5m for a 70m by 23m lounge\cite{indoorlocalization2}. We assume a higher localization error during the simulation for the purpose of robustness. We further assume people hold the mobile phone at random place at x,y plane between z = 0.5m and z = 2m. Therefore UE is located randomly in the x,y plane with z-coordinates between 0.5m and 2m, we call this area the \emph{test area}. Some initial measurements at the sample nodes are taken and the sorted beam selection map are generated by BIA. The number of sample nodes are about $1\%$ of the total number of grid nodes in the test area for both environments: 300 samples for condo, 640 samples for the office area. The sample nodes are uniformly distributed in the test area. We set $\xi_{BS}$ and $\xi_{UE}$ to 1 for both environments and $\beta = 5,20$ for one bedroom condo and office area. We simulate a new UE at 1000 randomly locations in the test area, and OBP is used to find the optimal beam selection. Fig. \ref{fig:cdf1},\ref{fig:cdf2} show the CDFs of number of entries of the sorted beam selection map tried before the ground truth beam selection is found with different localization accuracy. Assume that $BS\_ID_{v}'$,$BS\_Sec\_ID_{v}'$,$UE\_Sec\_ID_{v}'$ is the BS ID, BS transmit sector ID and UE transmit sector ID specified by the entry of $B^{sorted}_{v}$, by ground truth beam selection is found, we mean that $BS\_ID_{v}' = BS\_ID_{v}^{*}$, $BS\_Sec\_ID_{v}^{*}\in [BS\_Sec\_ID_{v}'-\xi_{BS},BS\_Sec\_ID_{v}'-\xi_{BS}]$ and $[UE\_Sec\_ID_{v}'-\xi_{UE},UE\_Sec\_ID_{v}'-\xi_{UE}]$, where $BS\_ID_{v}^{*}$, $BS\_Sec\_ID_{v}^{*}$ and $UE\_Sec\_ID_{v}'$ are the optimal BS ID, BS transmit sector ID and UE transmit sector ID.

The localization accuracy is interpreted in terms of the maximum localization error. Given the ground truth coordinate of UE to be $(x,y,z)$ and the maximum localization error equals $\delta$, the coordinates reported by the localization system is a uniform random variable distributed in a ball centered at $(x,y,z)$ with radius $\delta$. In order to analyze the inference accuracy of CRF we set $P_{TH}$ for OPB to a very small number such that all the entries in the sorted beam selection map will be tried. We notice that for one bedroom condo, the ground truth beam selection can be found within 6 trials more than $95\%$ of the times for all different localization accuracies. For office area, the ground truth beam selection can be found within 8 trials $98\%$ of the times for all different localization accuracies. The reason is that there are more BSs in the office area than condo, and more options for each UE to make connection, therefore it is harder to find the ground truth beam selection. We also notice that all the CDF curves does not reach 1, which indicates the ground truth beam selection can not be found within small number of trials, and traditional beamforming scheme should be used to find the ground truth beam selection. But this happens rarely ($\leq 5\%$ for condo, $\leq 2\%$ for office area). We also noticed that as the localization error decreasing, the ground truth beam selection can be found with less number of trials. This is because with higher localization accuracy, the location estimate is closer to the actual coordinates of the UE, and the inference result will be more accurate.

To compare the performance of the OBP with the traditional protocol in 802.11ad, we calculate the total amount of time taken for each UE to find its ground truth beam selection.  
With the traditional scheme, for a new UE entering the space, the BSs will broadcast beacons towards every direction periodically while UE receives with a quasi-omnidirectional pattern. The default beacon period is normally 100ms. And SLS and BRP will perform to set up the communication. Ignore the operation time of BS and UE and time taken for feedback frame, the total time taken for this beam forming procedure will be $T_{beacon} + T_{frame}\times 120 + T_{BRP}$, where $T_{frame}$ is the time taken to transmit a beacon, $T_{beacon}$ is the beacon period, $T_{BRP}$ is the time taken for BRP, and 120 is the total number of beacons BS and UE transmit since BS and UE each has 60 sectors. With InferBeam, the localization system will first locate the new UE, the best BS will transmit a beacon and wait another $T_{frame}$ for the reply. The total time taken equals $T_{loc} + N_{trials}\times 2T_{frame} + T_{BRP}$, where $T_{loc}$ is the localization latency, $N_{trials}$ is the number of trials before the optimal beam selection is found. Given $N_{trials}$ is normally within 6 for condo and 8 for the office area, and $T_{loc}$ is normally less than 100ms \cite{neurallocalization} \cite{FILA}\cite{{indoorlocalization1}}. InferBeam can reduce the process time by at least $(120-2N_{trials})T_{frame}$, which is around $86\%$ for office area and $90\%$ for condo. 

Fig. \ref{fig:additionalsamples} shows the number of samples needed for different number of transmit sectors of BS and UE, while keeping the $80\%$ prediction accuracy on the first trial and fixing the localization error (1m for condo, 4m for office area). We notice that the number of samples decreases as the number of sectors decreases. A reason is that with the decrease of the number of sectors, there are less options for each UE to make connection, therefore less samples are required to find the ground truth beam selection.
\subsection{Latency for beam adjustment}
To test the accuracy of the OBAP. We select 500 random points in the test area, finding their ground truth beam selections by using OBP. We simulate human blocking by placing a new obstacle with dimension $0.11m\times0.5m\times 1.7m$ on the floor with random locations in the x,y plane. Brick is chosen because the penetration losses of mm-wave through brick and human body are similar \cite{mmmeasure6}. We record the points whose ground truth beam selection changes due to this obstacle and perform OBAP on these points. We repeat this process for 100 times. Fig.~\ref{fig:additionalsamples} shows the number of additional trials required before the next best beam selection is found. Over $80\%$ of the points can find their next best beam with only 1 additional trial and over $90\%$ of the points can find their next best beam within 4 trials in both environments. 

The results show that InferBeam can greatly reduce the latency for beam adjustment. By using traditional scheme, the beamforming procedure defined above has to be re-triggered. For InferBeam the total time taken is only $N_{trials}\times 2T_{frame} + T_{BRP}$, where $N_{trials}$ is around 4 for all the environments. InferBeam can reduce the beam adjustment latency by more than $90\%$.
%
\section{Conclusion}
InferBeam introduced in this paper is a new protocol for
inferring best beam selection for a UE based on its location.
Based on a conditional random field (CRF) inference model,
InferBeam can prepare a list of best beam candidates in the
background for each location in the environment. Whenever
a UE determines it is necessary to switch to a new beam
(e.g., because the current beam is intermittently blocked by
another human being), the UE can quickly acquire a new beam
from its candidate list. As shown in the paper, just a few
around of testing these candidates would be sufficient. The
switching latency is under millisecond. This latency is orders
of magnitude lower than traditional beam sweeping methods.

We have especially designed the CRF model for the beam
alignment problem, in the choice of its parameters. As shown
in paper, learned parameters are insensitive to noise in the
training data. Thus, we can expect that CRF model is highly
transferable to different in-door environments. This is demonstrated
in our test scenarios where the test condos can have
different layouts from training ones.

In adapting to a new environment such as a condo unit,
InferBeam requires just a small amount of sampling. We have
show that the system can make best beam selection for $98\%$ of
locations, while sampling fewer than $1\%$ of locations. To the
best of our knowledge, the application of CRF to mm-wave
beam selection is new. We feel that it is a good approach,
given the CRF can leverage inherent consistency in best beam
selection among neighboring locations.

\appendix
\subsection{Proof of Theorem 1}
Remove all the terms which do not contain $w_{k}$ from $(8)$, $(8)$ can be written as 
\begin{multline}
\underbrace{-\log Z(\theta)}_{Part A}$ +  $\underbrace{\frac{1}{R}\sum_{r=1}^{R}\sum_{v\in V}\sum_{k=0}^{K}w_{k}\sum_{\{s|d(s,v)=k\}} \mathbbm{1}_{x^{r}_{s}=x^{r}_{v}}}_{Part B} + \\ \underbrace{\frac{1}{R} \sum_{k=0}^{K} -\frac{(w_{k}-\mu_{w_{k}})^{2}}{2\sigma_{w_{k}}^{2}}}_{Part C} + Const 
\end{multline}
where $Const$ include the terms which do not contain $w_{k}$. Given $Z(\theta)$ is the normalization constant for $P(x_{V},\theta)$ and $\frac{\partial Part A}{\partial w_{k}} = -\frac{\partial \log Z(\theta)}{\partial w_{k}}$. The normalization constant $Z(\theta)$ has the following expression:
\begin{multline}
Z(\theta) = \sum_{x_{V}} exp(\sum_{v\in V}\sum_{k=0}^{K} w_{k} \smashoperator{\sum_{\substack{\{s|d(s,v)=k\}}}} \mathbbm{1}_{x^{*}_{s}=x_{v}}+\smashoperator{\sum_{(v,v')\in E}} m_{vv'}\mathbbm{1}_{x_{v}\neq x_{v'}})
\end{multline}
therefore we have
\begin{align*}
&-\frac{\partial Part A}{\partial w_{k}} =-\frac{1}{Z(\theta)}\frac{\partial Z(\theta)}{\partial w_{k}}= \\& -\sum_{x_{V}} (\sum_{v\in V}\sum_{\{s|d(s,v)=k\}} \mathbbm{1}_{x^{*}_{s}=x_{v}})\frac{1}{Z(\theta)}exp(\sum_{v\in V}\sum_{k=0}^{K} w_{k} \smashoperator{\sum_{\substack{\{s|d(s,v)=k\}}}} \mathbbm{1}_{x^{*}_{s}=x_{v}}\\&+\sum_{(v,v')\in E} m_{vv'}\mathbbm{1}_{x_{v}\neq x_{v'}}) 
= -E_{P(x_{V};\theta)}(\sum_{v\in V}\sum_{\{s|d(s,v)=k\}} \mathbbm{1}_{x^{*}_{s}=x_{v}}) \numberthis
\end{align*}
By taking derivative of Part B w.r.t $w_{k}$, we get:
\begin{align}
\frac{\partial Part B}{\partial w_{k}} &= \frac{1}{R}\sum_{r=1}^{R}\sum_{v\in V}\sum_{\{s|d(s,v)=k\}} \mathbbm{1}_{x^{r}_{s}=x^{r}_{v}} \\&= E_{D}(\sum_{v\in V} \sum_{\{s|d(s,v)=k\}} \mathbbm{1}_{x^{*}_{s}=x_{v}})
\end{align}
Furthermore, we have $\frac{\partial Part C}{\partial w_{k}} = \frac{\mu_{w_{k}}-{w}_{k}}{R\sigma_{w_{k}}^{2}}$. Summing the three parts gives first part of the theorem. 
To prove the second part of theorem 2, we can rewrite $(8)$ as $\underbrace{-\log Z(\theta)}_{Part A}$ +  $\underbrace{\frac{1}{R}\sum_{r=1}^{R} \sum_{(v,v')\in E} m_{vv'} \mathbbm{1}_{x^{r}_{v}\neq x^{r}_{v'}}}_{Part B'} + \underbrace{\frac{1}{R} \sum_{k=0}^{K} -\frac{(m_{vv'}-\mu_{m})^{2}}{2\sigma_{m}^{2}}}_{Part C'} + Const$, where $Const$ are the terms which do not include $m_{vv'}$. By carrying out the similar step as above, we get $\frac{\partial Part A}{\partial m_{vv'}} = - E_{P(x_{V};\theta)} (\mathbbm{1}_{x_{v}\neq x_{v'}}), \frac{\partial Part B'}{\partial m_{vv'}} = E_{D}(\mathbbm{1}_{x_{v}\neq x_{v'}})$ and $\frac{\partial Part C'}{\partial m_{vv'}} =\frac{\mu_{m}-m_{vv'}}{R\sigma_{m}^{2}}$, summing them together gives the second part of the theorem.  $\square$
\subsection{Proof of Theorem 2}
We show the optimality of $\theta^{*}$ by proving $P(\theta|D)$ is a concave function, therefore the global maximum can be found by using gradient ascent method described in Algorithm 2. For the ease of interpretation, define $u_{k}=\sum_{v\in V}\sum_{\substack{\{s|d(s,v)=k\}}}\mathbbm{1}_{x^{*}_{s}=x_{v}}$, and $h = \sum_{v\in V}\sum_{k=0}^{K} w_{k} \sum_{\substack{\{s|d(s,v)=k\}}} \mathbbm{1}_{x^{*}_{s}=x_{v}}+\sum_{(v,v')\in E} m_{vv'}\mathbbm{1}_{x_{v}\neq x_{v'}}$. We use the same definition for Part A, Part B, Part C, Part B', Part C' 
as the proof of theorem 2, we then have 
\begin{equation}
\frac{\partial^{2} part C}{\partial w_{k} \partial w_{k'}} = \begin{cases}
-\frac{1}{R\sigma_{w_{k}}^{2}} &\text{$k'=k$}\\
0 &\text{otherwise}
\end{cases}
\end{equation}
\begin{equation}
\frac{\partial^{2} part B}{\partial w_{k} \partial w_{k'}} = 0
\end{equation}
\begingroup\makeatletter\def\f@size{9}\check@mathfonts
\begin{multline} 
\frac{\partial^{2} part A}{\partial w_{k} \partial w_{k'}} = -\sum_{x_{V}} u_{k}u_{k'} \frac{exp(h)}{Z(\theta)} -  \sum_{x_{V}} u_{k}\frac{exp(h)}{Z(\theta)} \sum_{x_{V}} u_{k'}\frac{exp(h)}{Z(\theta)} =\\ - E_{p(x_{V};\theta)} (u_{k} u_{k'})-E_{p(x_{V};\theta)}(u_{k}) E_{p(x_{V};\theta)}(u_{k'}) = - Cov(u_{k},u_{k'})
\end{multline} 
\endgroup
Similarly, we have $\frac{\partial^{2} part C}{\partial w_{k}\partial m_{vv'}} = \frac{\partial^{2} part B}{\partial w_{k}\partial m_{vv'}} = 0$, and $\frac{\partial^{2} part A}{\partial w_{k}\partial m_{vv'}} = -Cov(u_{k},\mathbbm{1}_{x_{v}\neq x_{v'}})$. 
We also have $\frac{\partial^{2} part C'}{\partial m_{vv'} \partial m_{qq'}} =  -\frac{1}{R\sigma_{m}^{2}}$ if $v=q,v'=q'$ and $0$ otherwise. $\frac{\partial^{2} part B'}{\partial m_{vv'} \partial m_{qq'}} = 0$, $\frac{\partial^{2} part A}{\partial m_{vv'} \partial m_{qq'}} = -Cov(\mathbbm{1}_{x_{v}\neq x_{v'}},\mathbbm{1}_{x_{q}\neq x_{q'}})$, $\frac{\partial^{2} part C'}{\partial m_{vv'} \partial w_{k}} = \frac{\partial^{2} part B'}{\partial m_{vv'} \partial w_{k}} = 0$, and $\frac{\partial^{2} part A}{\partial m_{vv'} \partial w_{k}} = -Cov(\mathbbm{1}_{x_{v}\neq x_{v'}}, u_{k})$. 
After deriving the expressions for all the second derivatives, we have $\frac{\partial^{2} P(\theta|D)}{\partial^{2} \theta} = - Cov(\vec{u},\vec{\mathbbm{1}}_{x_{v}\neq x_{v'}}) + \Phi$, where $\vec{u} = [u_{1},...,u_{K}]^\top$ and $\vec{\mathbbm{1}}_{x_{v}\neq x_{v'}} = [\mathbbm{1}_{x_{v}\neq x_{v'}}]^\top$, $(v,v')\in E$. $\Phi$ is the diagonal matrix which has negative constants $-\frac{1}{R\sigma_{w_{1}}^{2}},1\leq k \leq K$ and $-\frac{1}{R\sigma_{m}^{2}},(v,v')\in E$ on the diagonal. Therefore $\Phi$ is a negative semidefinite matrix, and $- Cov(\vec{u},\vec{\mathbbm{1}}_{x_{v}\neq x_{v'}})$ is also negative semidefinite matrix by definition. Therefore $\frac{\partial^{2} P(\theta|D)}{\partial^{2} \theta}$ is a negative semidefinite matrix and $P(\theta|D)$ is a concave function of $\theta$. $\square$
\subsection{Proof of Theorem 3}
The average received from $BS_{k}$, $E[P_{r}^{k}]$, equals E[$P_{r}^{k}|$blocked]P(blocked) + E[$P_{r}^{k}|$not blocked]P(not blocked), where E[$P_{r}^{k}|$blocked] is average received power given the LOS path is blocked. From \cite{mmmeasure1}, we have 
\begin{multline}
E[P_{r}^{k}|blocked]= 10\log_{10}(P_{t}G_{t}G_{r})-(PL_{FS}(d_{0})+\\29.5\log_{10}kd + X^{NLOS}_{\delta_{k}})
\end{multline}
and 
\begin{multline}
E[P_{r}^{k}|\text{not blocked}] = 10\log_{10}(P_{t}G_{t}G_{r}) -(PL_{FS}(d_{0})+\\12\log_{10}kd + X^{LOS}_{\delta_{k}}) 
\end{multline}
Given that P(not blocked) $= P_{e}kd$ and P(blocked) $= 1-P_{e}kd$, we have
\begin{multline}
E[P_{r}^{k}]=P_{e}kd(10\log_{10}(P_{t}G_{t}G_{r})-(PL_{FS}(d_{0})+29.5\log_{10}kd \\+ X^{NLOS}_{\delta_{k}})) + (1-P_{e}kd)(10\log_{10}(P_{t}G_{t}G_{r})-(PL_{FS}(d_{0})+\\12\log_{10}kd + X^{LOS}_{\delta_{k}}))= 10\log_{10}(P_{t}G_{t}G_{r})-PL_{FS}(d_{0}) - \\(17.5P_{e}kd+12)\log_{10}(kd)-(X^{NLOS}_{\delta}-X^{LOS}_{\delta_{k}})P_{e}kd-X^{LOS}_{\delta_{k}}
\end{multline}    
$\square$  
\subsection{Proof of Theorem 4}
First, we calculate $P(E[P_{r}^{k}]>E[P_{r}^{k'}])$, the probability of expected average received power from $BS_{k}$ is greater that from $BS_{k'}$:
\begin{multline}
P(E[P_{r}^{k}]>E[P_{r}^{k'}]) =\\ P(10\log_{10}(P_{t}G_{t}G_{r})-PL_{FS}(d_{0})-(17.5P_{e}kd+12)\log_{10}(kd)\\-(X^{NLOS}_{\delta_{k}}-X^{LOS}_{\delta_{k}})P_{e}kd-X^{LOS}_{\delta_{k}}>10\log_{10}(P_{t}G_{t}G_{r})-PL_{FS}(d_{0})\\-(17.5P_{e}k'd+12)\log_{10}(k'd)-(X^{NLOS}_{\delta_{k'}}-X^{LOS}_{\delta_{k'}})P_{e}k'd-X^{LOS}_{\delta_{k'}}) \\=P(X^{LOS}_{\delta_{k'}}-X^{LOS}_{\delta_{k}}+P_{e}d(k'X^{NLOS}_{\delta_{k'}}- k'X^{LOS}_{\delta_{k'}}-kX^{NLOS}_{\delta_{k}}+\\ kX^{LOS}_{\delta_{k}})>17.5P_{e}d(k\log_{10}(kd)-k'\log_{10}(k'd))+\\12\log_{10}(kd)-12\log_{10}(k'd))
\end{multline}
Ideally, to make prediction on every location in the space, $d$ should approaches 0. Moreover, $P_e$ is usually very small $(\approx 0.0078)$ \cite{blockmodel}, hence we have $P_{e}d \approx 0$, and we get:
\begin{multline}
P(E[P_{r}^{k}]>E[P_{r}^{k'}]) = P(X^{LOS}_{\delta_{k'}}-X^{LOS}_{\delta_{k}}>12\log_{10}(kd)-\\12\log_{10}(k'd))=
P(X^{LOS}_{\delta_{k'}}-X^{LOS}_{\delta_{k}}>12\log_{10}(\frac{k}{k'})
\end{multline}
Since $X^{LOS}_{\delta_{k'}},X^{LOS}_{\delta_{k}}\sim \mathcal{N}(0,1.8)$, we have $X^{LOS}_{\delta_{k'}}-X^{LOS}_{\delta_{k}}\sim \mathcal{N}(0,1.8\sqrt{2})$, and $P(E[P_{r}^{k}]>E[P_{r}^{k'}]) = \mathcal{Q}(\frac{12\log_{10}(\frac{k}{k'})}{1.8\sqrt{2}})$.
To calculate $Q(x_{v_{0}}=Tx_{k})$, we calculate $P(E[P_{r}^{k}]>E[P_{r}^{k'}])$ for every $1\leq k'\leq K, k'\neq k$. We have $Q(x_{v_{0}}=Tx_{k}) = \prod_{k'=1,k'\neq k}^{K} \mathcal{Q}(\frac{12\log_{10}(\frac{k}{k'})}{1.8\sqrt{2}})$. Substituting $Q(x_{v_{0}}=Tx_{k})$ to $(15)$, we get $w_{k} = \log(Q(x_{v_{0}}=Tx_{k})) + C = \sum_{k'=1,k'\neq k}^{K}\log(\mathcal{Q}(\frac{12log_{10}(\frac{k}{k'})}{1.8\sqrt{2}})) + C$. $\square$
\end{document}